\documentclass[structabstract]{aa}

\usepackage{epsfig,graphicx,latexsym,amsmath,amssymb}
\usepackage{natbib}
\usepackage{mathrsfs}
\usepackage{lastpage}
\usepackage{color}

\bibpunct[,]{(}{)}{;}{a}{}{,}

\def\lsim{\mathrel{\rlap{\lower 3pt\hbox{$\sim$}}\raise 2.0pt\hbox{$<$}}}
\def\gsim{\mathrel{\rlap{\lower 3pt\hbox{$\sim$}} \raise 2.0pt\hbox{$>$}}}
\usepackage{txfonts}
\begin{document}

   \title{Retrograde binaries of massive black holes in circum-binary accretion discs}

   \subtitle{}

   \author{Pau Amaro-Seoane
          \inst{1}
          \and
          Cristi{\'a}n Maureira-Fredes\inst{1}          \and
          Massimo Dotti
          \inst{2,\,3}
          \and
          Monica Colpi
          \inst{2,\,3}
          }

   \institute{Max Planck Institut f\"ur Gravitationsphysik
              (Albert-Einstein-Institut), D-14476 Potsdam, Germany\\
              \email{Pau.Amaro.Seoane@gmail.com, Cristian.Maureira.Fredes@aei.mpg.de}
         \and
             Universit{\`a} di Milano Bicocca, Dipartimento di Fisica,
      G. Occhialini, Piazza della Scienza 3,
      I-20126, Milano, Italy\\
             \email{Massimo.Dotti@mib.infn.it, Monica.Colpi@mib.infn.it}
         \and
            INFN, Sezione di Milano-Bicocca, Piazza della Scienza 3,
      I-20126, Milano, Italy
             }

   \date{\today}

  \abstract
{
We explore the hardening of a massive black hole binary embedded in a
circum-binary gas disc, under a specific circumstance: when the binary and the
gas are coplanar and the gas is {\it counter-rotating}. The binary has unequal
mass and the interaction of the gas with the lighter secondary black hole is
the main cause of the braking torque on the binary that shrinks with time.  The
secondary black hole, revolving in the direction opposite to the gas,
experiences a drag from gas-dynamical friction and from direct accretion of
part of it.
}
{
In this paper, using two-dimensional (2D) hydrodynamical grid
simulations we investigate the effect of changing the accretion prescriptions
on the dynamics of the secondary black hole which in turn affect the binary
hardening and eccentricity evolution.
}
{
We find that realistic accretion
prescriptions lead to results that differ from those inferred assuming
accretion of all the gas within the Roche Lobe of the secondary black hole.
}
{
When considering  gas accretion within the gravitational influence radius of
the secondary black hole (which is smaller than the Roche Lobe radius) to
better describe gas inflows,  the shrinking of the binary is slower.  In
addition, in this case, a smaller amount of accreted mass is required to reduce
the binary separation by the same amount.  Different accretion prescriptions result in different disc's surface densities
which alter the black hole's dynamics back.
Full 3D SPH realizations of a number of representative
cases, run over a shorter interval of time, validate the general trends
observed in the less computationally demanding 2D simulations.

}
{
Initially circular black hole binaries
increase only slightly their eccentricity which then oscillates around small values
($<0.1$) while they harden. By contrast, initially eccentric binaries become more and more eccentric.
  A
semi-analytical model describing the black hole's dynamics under accretion only
explores the late evolution stages of the binary in an otherwise unperturbed
retrograde disc to illustrate how eccentricity evolves with time in relation to
the shape of the underlying surface density distribution.
}
   \keywords{giant planet formation --
                $\kappa$-mechanism --
                stability of gas spheres
               }
   \maketitle

\section{Introduction}
\label{sec.intro}

Massive black hole pairs are thought to be the natural outcome of galaxy
mergers along the cosmic history \citep{BBR80}.  When two galaxies collide, the
gravitational interaction of their galactic cores with the underlying dark
matter, stellar and gaseous background guides the sinking of the two massive black
holes (MBH) at the center of the galaxy remnant, leading to the formation of a
Keplerian binary. This occurs when the mass in gas and stars enclosed within
the MBH orbit is smaller than the masses of the two MBHs, typically of $\sim$
parsec for MBH masses of million suns \citep{Colpi14}.  If the binary
further hardens to attain separations as small as  $\lsim 0.001$
pc\footnote{The exact separation at which a MBH binary coalesces in less
than an Hubble time depends on the binary eccentricity, mass, and mass ratio
\citep[see e.g.][for an approximation based on Keplerian ellipses]{Peters64}.},
the emission of gravitational waves forces the two MBHs to coalesce in less
than an Hubble time. Their final pairing and coalescence will be measurable with a LISA-like observatory like eLISA if the
total mass is $\lessapprox 10^7\,M_{\odot}$
\citep{Amaro-SeoaneEtAl2012,Amaro-SeoaneEtAl2012b}.

The evolution of the massive MBH binary (MBHB hereon) on sub-pc scales depends
on the properties of the cores of their hosts. In gas-poor remnants, MBHBs lose
energy and angular momentum via scattering individual stars. The final fate of
the binary depends on the effective reservoir of stars the MBHs can interact
with.  It has been shown that the presence of some degree of
tri-axiality~\citep[naturally present in galaxy merger
remnants,][]{PretoEtAl2011,KhanEtAl2011,KhanEtAl12} can excite centrophilic
orbits in a large fraction of the stars of the galaxy remnant, bringing the two
MBHs to the scales of gravitational wave driven inspiral and coalescence
\citep[see also][]{BerczikEtAl06,
BerentzenEtAl09,GualandrisMerritt2012,KhanHolley-Bockelmann2013}.

The presence of dense gas structures could accelerate the evolution of the
binary, resulting in a faster coalescence~\citep{BBR80}.  A fundamental
difference between gas-poor and gas-rich environments is that, in presence of a
consistent amount of gas, accretion onto the MBHs is expected to be important.
This would allow us to detect dual AGN as
spatially resolved nuclear sources, and MBHBs at shorter separations from
peculiar signatures in their optical/X-ray spectra~\citep[e.g.][]{BBR80,
ShenLoeb10, Tsalmantza11, Eracleous11, Montuori11, Montuori12,
Sesana12, Tanaka12, Dorazio12,Bogdanovic15,Dorazio2015}, allowing us to map the orbital decay of MBHs in
the electromagnetic spectrum.  In gas-poor environments, the most robust way of
identifying MBHBs is through the detection of gravitational radiation during
the inspiral phase. For large MBH $\gsim 10^8\, M_\odot$, the Pulsar Timing
Array experiment, operating at nano-Hz  frequencies, might reveal their signal
\citep{HobbsEtAl2010}. Furthermore, MBHBs can be detected during the inspiral,
merger and ring-down in experiments such as eLISA, at shorter wavelengths
(around 0.1 mHz-1 Hz), and for lighter MBHB coalescences ($\sim 10^7\,M_\odot$)
\citep{Amaro-SeoaneEtAl2012}.

The presence of massive gas structures close to MBHBs is not
unexpected. If the two merging galaxies initially had some significant
amount of gas, their reciprocal perturbation
drives gas inflows toward the center of the two structures. These
inflows result in massive gas discs in the center of the
galaxy remnant~\citep[e.g.][]{Escala05,mayer07,Dotti07,hopkins10,Dotti12}. However, the details
of the interaction between the binary and a gas disc are still
debated \citep{fiacconi13,roskar15,lupi15,delvalle2015}. The gas disc can be either corotating or counter-rotating with
respect to the MBHB~\citep{nixon11a,roedig14}. The corotating case seems to be
the more natural outcome of a gas rich galaxy merger~\citep[see,
  e.g.][]{mayer07}, since the MBHs bind in a binary during the natal
process that forms the nuclear gas disc. For this reason, MBHBs
embedded in co-rotating circum-binary discs have been extensively
studied~\citep{Goldreich80, Lin86, Artymowicz94, Ivanov99, Escala05,
  Hayasaki07, Dotti07, Dotti09, M&M08, Cuadra09, Lodato09, Farris11,
  Roedig11, Roedig12, Noble12, Dorazio12}. However, the degree of
misalignment between the gas in the remnant nucleus and the binary
could depend on the parameters of the merger~\citep{Blecha11,
  Hopkins12} so that counter-rotating accretion is not ruled
out. Furthermore, if the binary does not coalesce on a short timescale
(comparable with the timescale over which star formation depletes the
central co-rotating gas), subsequent inflows of gas could be
uncorrelated to the angular momentum of the binary, possibly resulting
in counter-rotating circum-binary discs.

Regarding the evolution of a MBH binary, retrograde and prograde discs
differ on a few important aspects: (i)
in the prograde scenario the disc-binary interaction leads to
 the opening of a gap, i.e. a hollow region surrounding the MBHs
 of size comparable to twice the binary separation, and binary hardening
 is somewhat reminiscent to planet type II migration.
In the retrograde scenario resonances are either absent (for circular
binaries) or weak \citep{NixonLubow2015}. In many cases the
binary does not manage to excavate a gap \citep{Bankert15}, and the MBHB-disc interaction
actually enhances the inflow of matter toward the center~\citep{nixon2011b};
(ii) retrograde gas
interacting with the MBHs can remove more angular momentum per unit of
mass than in the prograde case, since its initial angular momentum has
sign opposite to that of the binary and this leads to an increase in the eccentricity \citep{nixon2011b,roedig14,Schnittman15}; (iii) in the retrograde case, the
relative velocities between the MBHs (in particular between the
secondary MBH, in an unequal mass binary) and the disc are
significantly larger, so that the interaction between the gas and the
MBHs is confined to smaller regions.

In this work we present a suite of 2D hydro-dynamic simulations to
study in detail the evolution of a MBHB embedded in a counter-rotating
disc with focus on how different prescriptions on the accretion onto the secondary MBH can influence
the  evolution of the orbital elements. In Section~\ref{sec.NumTool} and Section~\ref{sec.AccrPrescr}  we describe the
numerical tools,  the initial configurations and the accretion prescriptions. In \ref{sec.Results}, we evaluate the effect of the
different accretion recipes on the dynamics of the binary.   Furthermore, we present results obtained from a set of 3D SPH simulations
of MBHBs in a retrograde non-self gravitating disc to highlight
commonalities and differences with the results from 2D simulations.
In section~\ref{sec.Semianal}  we present a semi-analytical model in order to explore the long term evolution
of the binary. We discuss the
implications of our findings in Section \ref{sec.Discuss}.

\section{Numerical tool and description of the initial models}
\label{sec.NumTool}

We consider the case of MBH binaries orbiting in the orbital plane
of an accretion disc. This allows us to limit the dimensions in our
simulations to two. Thus, we can use {\sc
Fargo}\footnote{\url{http://fargo.in2p3.fr}}, a two-dimensional hydrodynamical
grid program that integrates the isothermal Navier-Stokes equations using a
staggered polar grid \citep{Masset00}.

The code is particularly suited for quasi-Keplerian scenarios, since it
separates the azimuthal averaged motions from azimuthal and radial
perturbations, resulting in longer time steps \citep[the FARGO algorithm was
originally presented in][]{Masset00}. This algorithm speeds up significantly the
calculations and hence allows us to study the parameter space more efficiently
than with other schemes. {\sc Fargo} is parallelised by splitting radially the
grid in rings, which are calculated in different CPUs.

\begin{figure}
\resizebox{\hsize}{!}
          {\includegraphics[scale=1,clip]{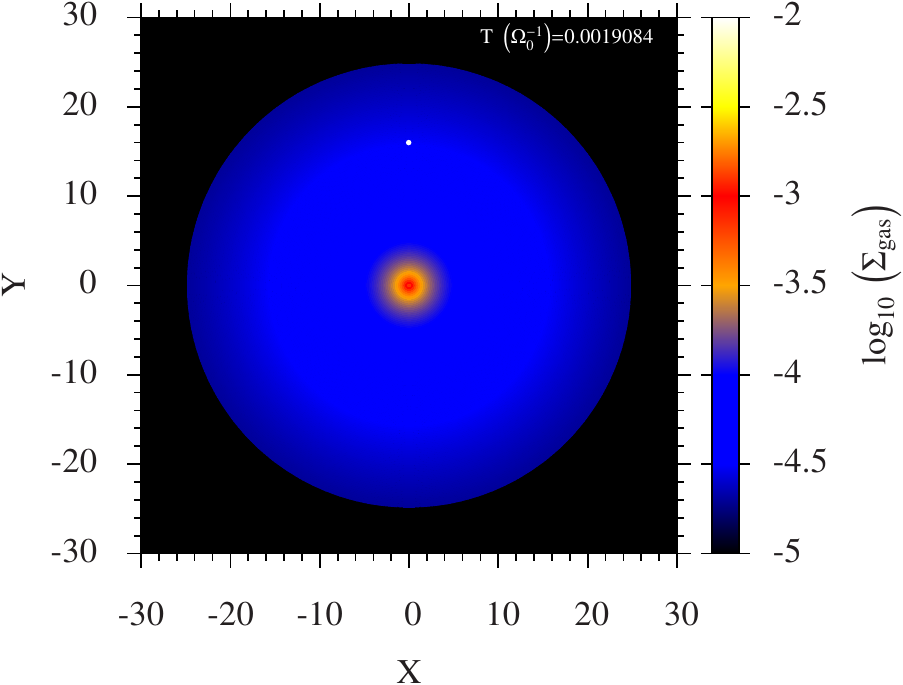}}
\caption
{
Face-on, colour coded map of the disc's surface density at the onset of
each simulation.
The position of the secondary is marked with a solid green circle.
For all quantities, we use the internal units of the code, except for the time,
which is in units of $\Omega_0^{-1}$, i.e. the inverse of the initial binary's
rotational frequency.
}
\label{fig.initialcond}
\end{figure}

We run a set of 14 simulations of unequal mass MBHBs  in gaseous discs.
In every run the primary MBH ($M_1$) is treated as an external potential, and
is not evolved during the simulations; it is considered a point-like source.
In internal units, the mass of $M_1$ is 1, and is placed at rest at the center
of the disc. The disc is confined within an outer radius defined by the radial
limits of the grid. In internal units, the outer radius is 25, and the inner
radius is 0.1. The disc follows a Mestel profile, as we depict in
Figure~\ref{fig.initialcond}, with a surface density
$\Sigma_{\rm gas}({\hat R}) \propto {\hat R}^{-1}$,
and a total mass of 0.078, i.e.  $\approx 1/12$ of the mass of the primary.

The disc follows an isothermal equation of state, with a thermal profile
resulting in an aspect ratio $H/{\hat R}=0.04$ constant throughout the disc.
The dynamics of the disc is initially quasi-Keplerian, since, globally, the
potential is dominated by $M_1$.  The initial angular speed is however not strictly
Keplerian, since the code accounts for the pressure support to the rotational
equilibrium.  The computational domain is divided in a grid of 128 radial and
384 azimuthal sectors.

In every simulation a second MBH, $M_2$, of initial mass $M_{2,0}=0.1\,M_1$ is
placed in the disc, at a distance $d=10$ from $M_1$. The secondary is
initially moving on a bound orbit, and counter-rotates with respect to the
disc. $M_2$ can either be on a circular or on an eccentric orbit. In this last
case we choose the initial eccentricity to be $e_0=0.6$, corresponding to an
initial apocenter ${\hat R}_{\rm apo} = 16$. At the beginning of each
simulation, the secondary MBH is implanted in the disc as a particle with mass
increasing from 0 to $M_{2,0}$ over one orbital timescale.  This choice
prevents the growth of unphysical disc perturbations close to the
secondary.

We pay particular attention on how the gas is added and removed from the disc.
The amount of gas present can change only if it crosses the radial edges of the
computational domain, or if it is accreted by the secondary. At the outer and
inner edge of the computational domain we set outflowing boundary conditions,
i.e. matter crossing the boundaries disappears from the computational domain,
but no gas can inflow into the computational domain.

So as to check the stability of the initial conditions, we first corroborate
that the disc is stable by setting the mass of the secondary to a very small
value (a 100th of the primary). We confirm the stability for what indeed is the
range of mass ratios characteristic to the code -written to study the migration of a planet of mass much smaller than that of the central star- for some tens of initial periods of the binary.

Every set of initial conditions has been run five times, using
different prescriptions for the accretion onto $M_2$. The secondary is
modelled either as a sink particle that can accrete gas from the
disc, or as a point mass whose mass is fixed in time. Since the
accretion of mass has strong consequences on the dynamical evolution
of $M_2$ we use different prescriptions for the mass accretion, as it
is discussed in the next section.

\section{Accretion prescriptions}
\label{sec.AccrPrescr}

Accretion onto the secondary MBH is a key process affecting the dynamical
evolution of the MBHB as (i) the accreting gas changes the mass and velocity
of $M_2$, according to conservation of the total momentum of the system, and
(ii) the process of accretion itself decreases the gas density close to the
secondary.  In addition, the perturbation induced by the motion of $M_2$
further changes the underlying disc density pattern,  back reacting on
the dynamics of the MBHs in the binary.  It is therefore important to implement
an accretion prescription that does not bias the evolution of the binary. The
safest approach would be to follow the hydrodynamics of the gas down to the
innermost stable circular orbit around each MBH, or at least around the
secondary MBH.  Our simulations however do not model all the physics needed to
evolve gas down to such extremely small radii.  Further out from this physical
limit, the numerical nature of our investigation prevents us to set a too
small value of the sink radius, that in a number of cases is a free parameter
of the simulation.  When gas reaches the sink radius, the gas itself is
removed from the computational domain and its mass and momentum is added to
the MBH. As a consequence, we decided to use different prescriptions for the
gas accretion onto the secondary, to test if any accretion prescription
results in an artificial orbital evolution of the secondary.

The first prescription we use is the standard FARGO implementation
that we will refer as RL model, hereon: gas is accreted onto the
secondary if its distance from $M_2$, denoted as $R^{\rm gas}_2,$ is
less than a given fraction (0.75) of the Roche Lobe (RL) radius
$R_{\rm RL}$ around $M_2$.  The mass accretion is modelled by reducing the gas
density within the RL by a factor $(1-f_{\rm red})$, at every timestep
$\Delta t$. To prevent spurious large density and pressure jumps,
$f_{\rm red}=1/3$ if $0.45 R_{\rm RL} < R^{\rm gas}_2 < 0.75 R_{\rm
  RL}$, while $f_{\rm red}=2/3$ if $R_2< 0.45 R_{\rm RL}$ \citep[see
  for more details][]{Kley99}.

However, we note that for a binary counter-rotating with respect to the
accretion disc in which it is embedded, the Roche Lobe of the secondary MBH is
far from being comparable to its gravitational sphere of influence radius
$R_{\rm bound}$. The counter-rotating gas crossing the Roche Lobe in the region
$R_{\rm bound} < R^{\rm gas}_2 <R_{\rm RL}$ is moving too fast with respect to the
secondary MBH to bind to it.  For $M_2 \ll M_1$, the Roche Lobe radius is

\begin{equation}
  R_{\rm RL} \sim \frac{1}{2}d\left(\frac{M_2}{M_1}   \right)^{1/3},
\end{equation}
where $d$ is the separation between the two MBHs, while
\begin{equation}
 R_{\rm bound} \sim \frac {G \, M_2}{V_{\rm rel}^2} \sim \frac {1}{4}d \frac {M_2}
{M_1},
\end{equation}
where $G$ is the gravitational constant, and $V_{\rm rel}$ is the modulus of the relative
velocity between the gas and the secondary. The ratio between the two
radii is
then
\begin{equation}
\frac{R_{\rm bound}}{R_{\rm RL}} \sim \frac{1}{2} \left(
\frac{M_2}{M_1}\right)^{2/3}.
\end{equation}
As a consequence, the Roche Lobe based standard implementation of accretion
could result in an overestimated accretion rate, hence in an unphysical
dynamical evolution of the secondary.

A second prescription we implement is based on the choice of a fixed
sink radius $R_{\rm fix}$.  To prevent spurious pressure jumps we use
the same two zones implementation discussed above, accreting 1/3 of
the material present in the $0.5 \,R_{\rm fix} \,< \,R^{\rm gas}_2\, <
\, R_{\rm fix}$ shell and 2/3 of the material with $R^{\rm gas}_2\, <
\, 0.5 \,R_{\rm fix}$ at each timestep.  The size of the
fixed sink radius around $M_2$ could affect the dynamics of $M_2$ in
an unphysical manner if $R_{\rm fix} > R_{\rm bound}$. To check for
this spurious effect we run the same set-up with different values of
$R_{\rm fix}=$0.5, 0.25, and 0.05, in code units.  For a circular
binary at the onset of the simulation $R_{\rm fix}/R_{\rm bound}$ is
approximatively 2, 1, and 0.2 for $R_{\rm fix}=$0.5, 0.25, and 0.05,
respectively; similarly $R_{\rm fix}=0.5$ corresponds to $R_{\rm
  fix}/R_{\rm RL} \approx 0.2$.

A third prescription we test requires gas to be gravitationally bound
to $M_2$ in order to get accreted. This automatically solves the problem of
unbound gas spuriously binding (and accreting) to the secondary. In this case,
the gas density reduction factor is either $f_{\rm red}=1/3$ if the total
 gas energy per unit mass (with respect to the secondary) is $ (3/4) W < E < (1/2) W$, or
 $f_{\rm red}=2/3$ if $E < (3/4) W$, where $W=-G M_2/R_2$.

Finally, we test the dynamical evolution of an accreting secondary against a
non-accreting one. In this case, the secondary MBH is allowed to bind gas
according to the above prescription, but the bound gas is not removed from the
simulation, and can either remain bound to $M_2$, co-moving with it on
retrograde orbits, or can be stripped by either the tidal field of the primary
or by the ram pressure of the gas disc.  In the following, for the standard
FARGO implementation we will refer to as the ``RL model'', as ``Fix'' 0.5,
0.25, and 0.05 for the models with fixed sink radii, and as ``Bound'' and ``No-accretion'' for the remaining last two.

\section{Results}
\label{sec.Results}

\subsection{Effect of the accretion prescription on the dynamics}\label{sec:acc}

\begin{figure}
\resizebox{\hsize}{!}
          {\includegraphics[scale=1,clip]{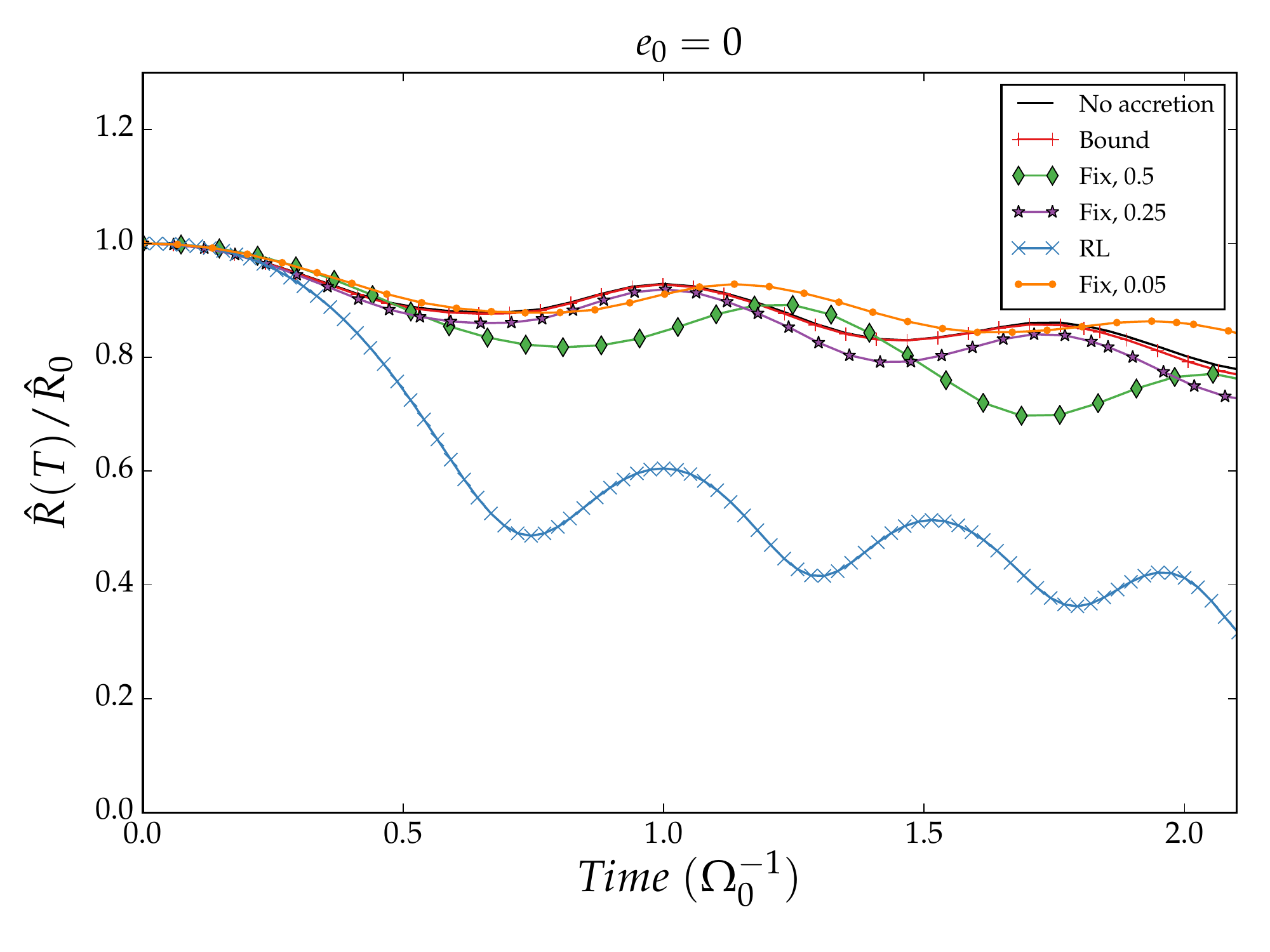}}
\caption
{
Relative MBH orbital separation (in units of the initial separation) as a
function of time in units of $\Omega_0^{-1}$ for a binary with initial
eccentricity $e_0=0$. Only the first $\sim 2$ orbits are shown to highlight
the differences in the orbital evolution caused by the different accretion
prescriptions. Black solid line refers to the run without any implementation
of accretion onto the secondary (the no-accretion model).
The red plus line refers to the Bound model and the blue crosses line
to the RL model (lowest line). Green diamonds, purple stars and orange circles refer
to the Fix models with sink radii $R_{\rm fix}=0.5$, $0.25$, and $0.05$,
respectively. Colour version is available in the on-line version.
}
\label{ashort}
\end{figure}

Figure~\ref{ashort} shows the evolution of the binary separation
as a function of time, for different accretion prescriptions over a short timescale
(i.e., an interval of 2 orbital times). Initially the MBHs move on a circular orbit.

The timescale of MBH evolution strongly depends on the accretion
prescription assumed. Over a timescale of a few orbital times,
the binary hardening in the RL model  is faster that in the
Bound model, as illustrated  in Figure~\ref{ashort}.
The reason for this is simple. Gas at a distance from $M_2$ in between the
influence radius of the secondary and its Roche Lobe ($R_{\rm bound} < R^{\rm gas}_2 <
R_{\rm RL}$) is, by definition, not bound.
When the sink radius is forced to be equal
to $R_{\rm RL}$, the full momentum of the gas
(i.e., also the momentum of unbound gas) is added to the MBH,
resulting in an artificially fast
dynamical evolution of the binary, because
of the unrealistically high accretion rate.

We can test the above interpretation exploring additional cases varying
$R_{\rm fix}$.  Whenever a fixed sink radius is assumed, the dynamics depends
on the size of the sink radius itself. As expected, larger sink radii
correspond to faster migration.  In Figure~\ref{ashort} we show how the
evolution of the MBH separation, which is sensitive to $R_{\rm fix}$, converges
with continuity to the Bound model when decreasing the values of the sink
radius. We further find that the No-accretion model in which the MBH does not
accrete results in a binary decay very similar to the Bound model. The reason
is straightforward: In both cases gas is allowed to bind to the secondary
MBH. In the No-accretion run, the gas transfers its (negative) linear momentum
to the secondary when it starts co-moving with it, i.e. during the binding
process. The same happens in the Bound model in which the gas either binds to
the MBH (transferring its linear momentum) or gets immediately accreted during
the binding process. In this last case the linear momentum conservation is
forced by the accretion prescription. Furthermore, material that would not
bind to the secondary is not accreted by default.
We further notice that in the Fix models, when the sink radius of $M_2$ is
larger than its gravitational influence radius, the MBH absorbs the linear
momentum of gas that, in reality, would not be fated to interact with it.
Under these conditions the MBHB separation again has a fast artificial decay.

\begin{figure}
\includegraphics[width=\hsize]{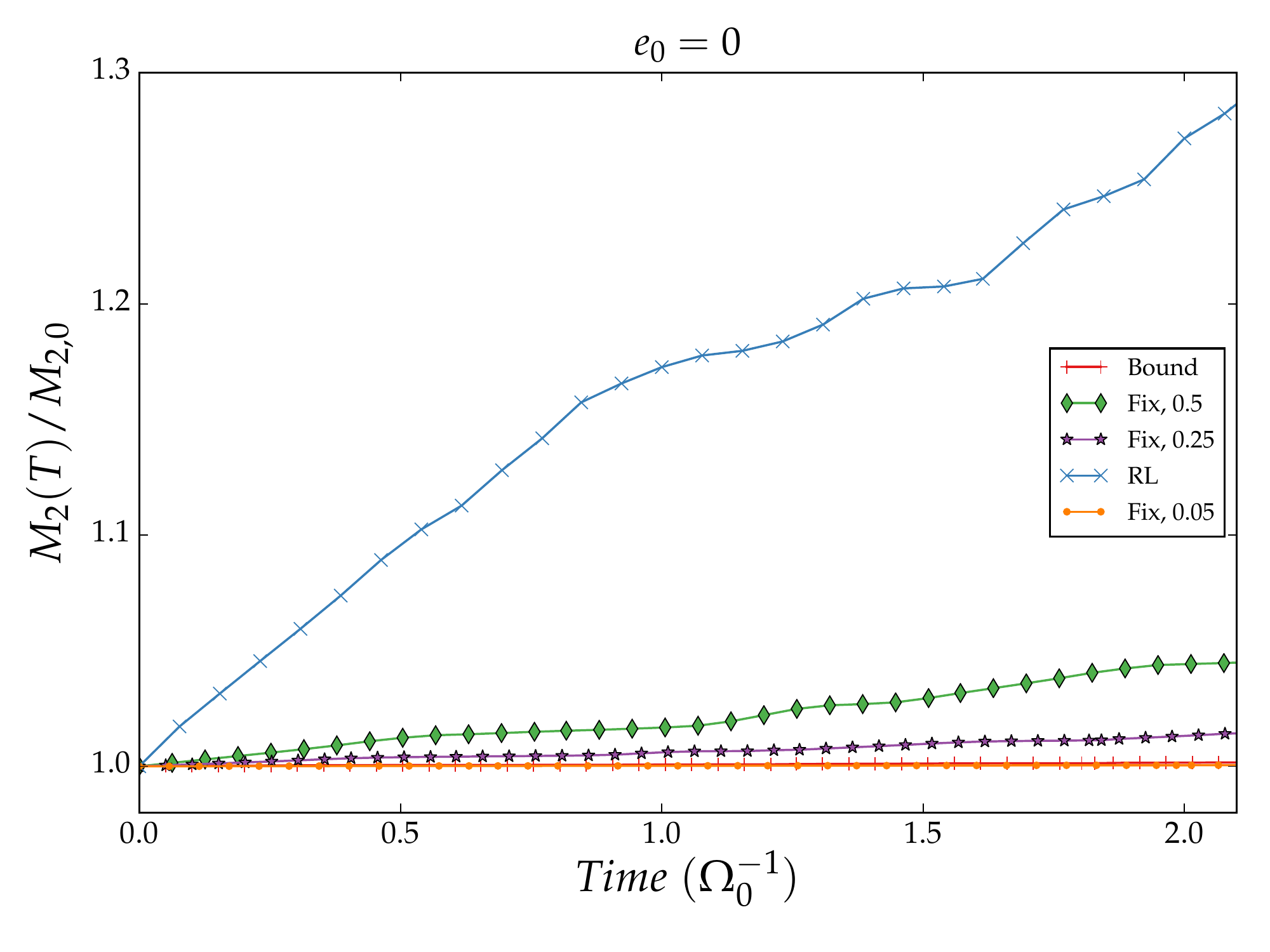}
\caption
 {
 Mass of the secondary MBH, in units of the initial mass $M_{2,0}$, as
 a function of time in units of $\Omega_0^{-1}$.
 Line colour and style codes are as in Figure~\ref{ashort}.
 }
\label{mshort}
\end{figure}

The amount of mass that is accreted by the secondary MBH can be used as a
tracer of the linear (and angular) momentum that the disc transfers to the MBH,
and in turn to the MBHB.  Figure~\ref{mshort} shows the evolution of the
secondary mass as a function of time, for the different accretion
prescriptions.  The MBH in the RL run displays the fastest mass growth and
largest amount of accreted mass. In particular in the RL model, the MBH
accretes 30\% of its mass in order to reduce the separation by 25\%. By
contrast a reduction of 25\% in the relative MBH distance is attained with a
fractional mass increase $\sim 0.3\%$ in the Bound model.  This indicates that
in the process of binary hardening, the fractional decay per unit accreted mass
is less effective in the RL model.  In this case, the large mass accreted by
the MBH creates an empty region resulting in a much weaker frictional drag on
the MBH from the torque of surrounding gas particles that contribute to the
deceleration without being accreted.  An interpretation of the results will
also be discussed in Section~\ref{sec.Semianal}.

\subsection{Evolution of circular binaries}

\begin{figure*}
    \centering
    \begin{minipage}{.45\textwidth}
      \centering
      \includegraphics[width=\textwidth,clip]{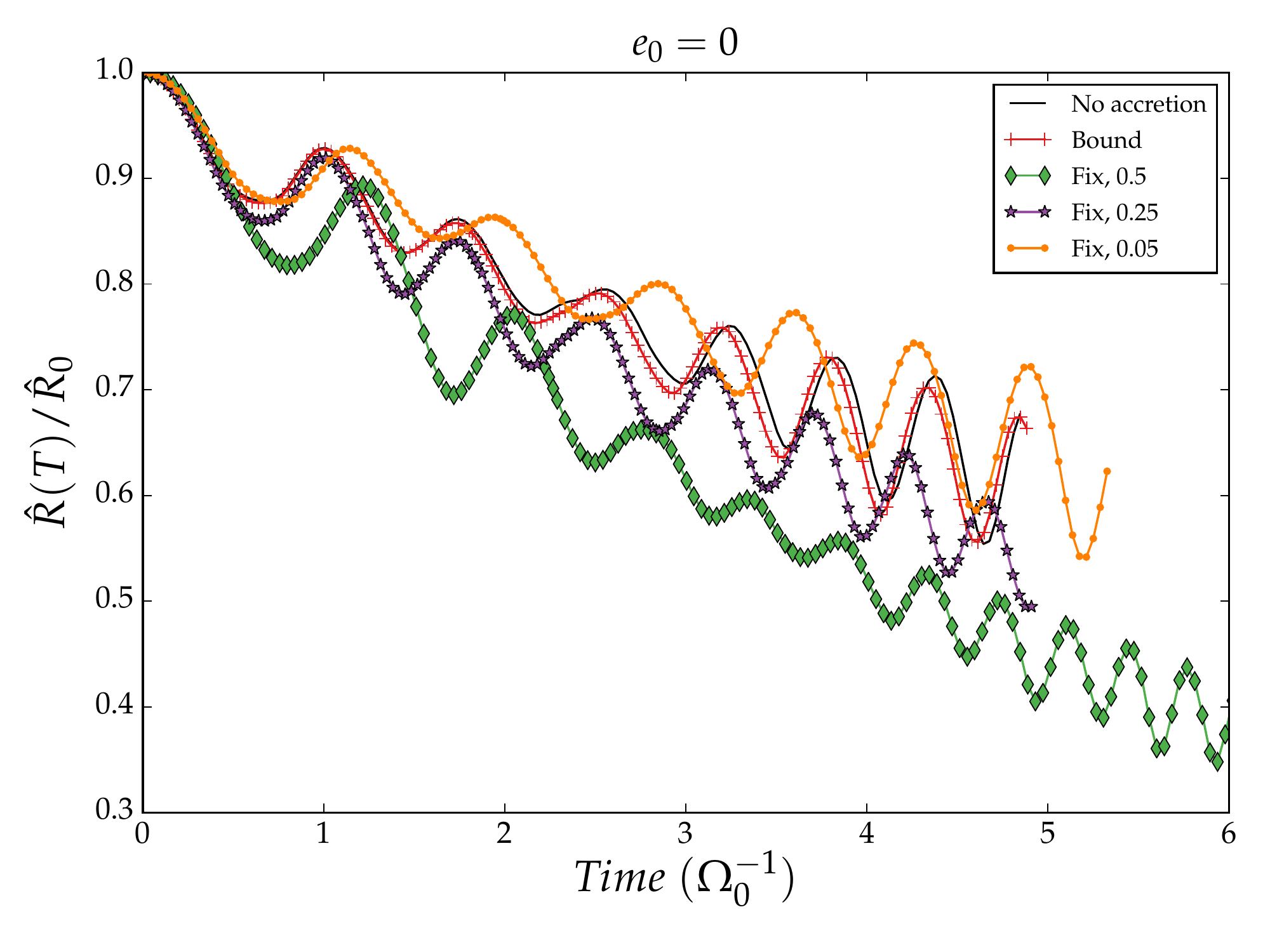}
    \end{minipage}
    \begin{minipage}{.45\textwidth}
      \centering
      \includegraphics[width=\textwidth,clip]{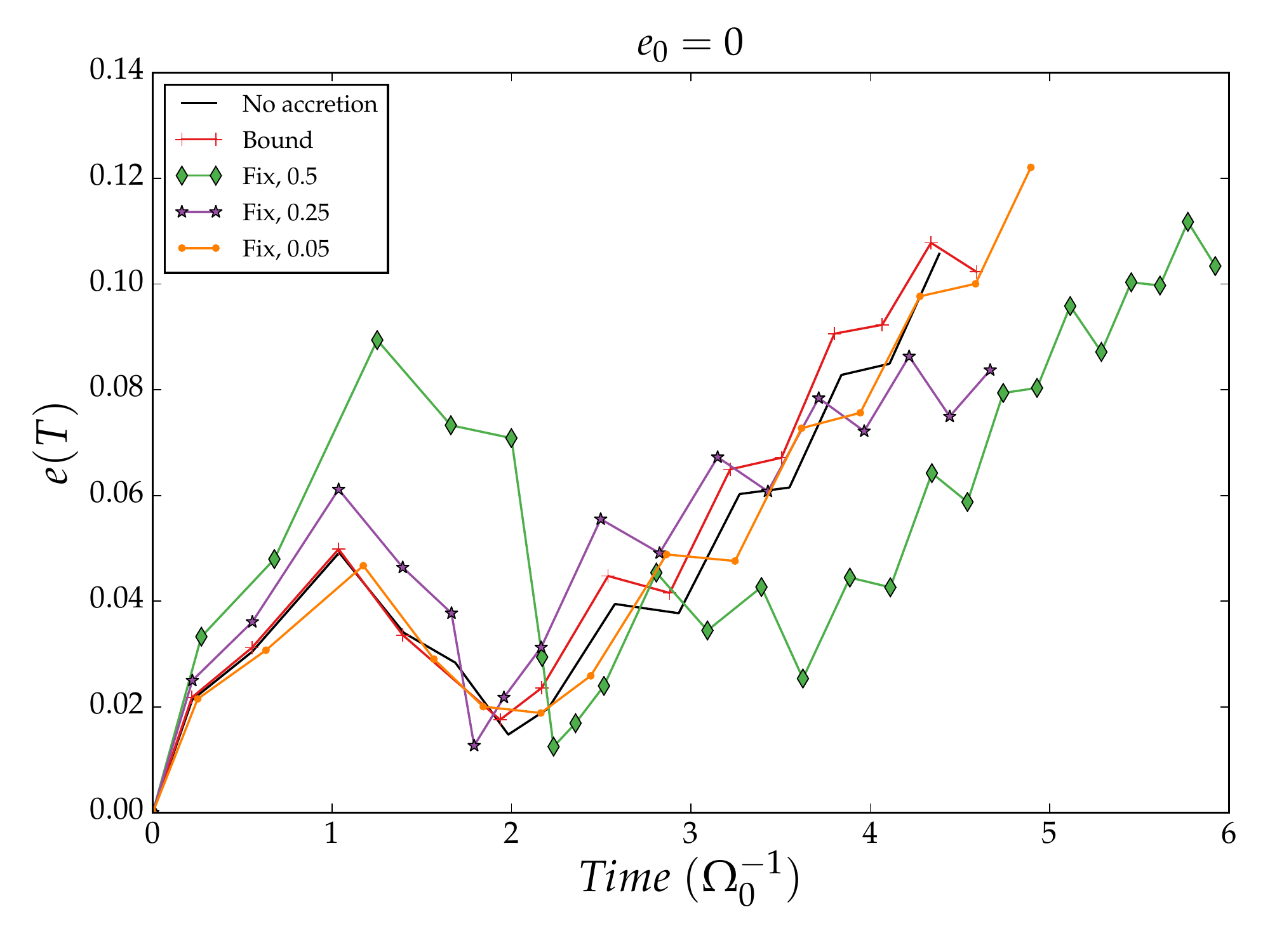}
    \end{minipage}
    \caption{
      MBHB orbital separation (left) and eccentricity  (right)  as a function of
      time, for $e_0=0$. We follow the same notation as in Figure~\ref{ashort}.
      The RL model is not considered, here.
    }
    \label{fig.Separation_Ecc_Time_Circular}
\end{figure*}

The evolution of a circular binary is studied further in this section.
Figure~\ref{fig.Separation_Ecc_Time_Circular} illustrates the run of the BH
separation versus time over $\sim 6\,\Omega_0^{-1}$, for the different
accretion prescriptions. The RL case is not included in the figure because of
the too large, unrealistic accretion prescription. On such a long timespan, all
the other runs but for the Fix 0.5 show similar evolutions. The discrepancy between
Fix 0.5 and the other runs can be ascribed to the large sink radius of Fix 0.5
resulting in an overestimate of the accretion rate and of the orbital brake.
As shown in Figure ~\ref{fig.Separation_Ecc_Time_Circular} the binary
eccentricity, increasing slightly at start, fluctuates around a mean of 0.08
and never exceeds $0.16$ during the whole evolution.

\begin{figure}
\includegraphics[width=\hsize]{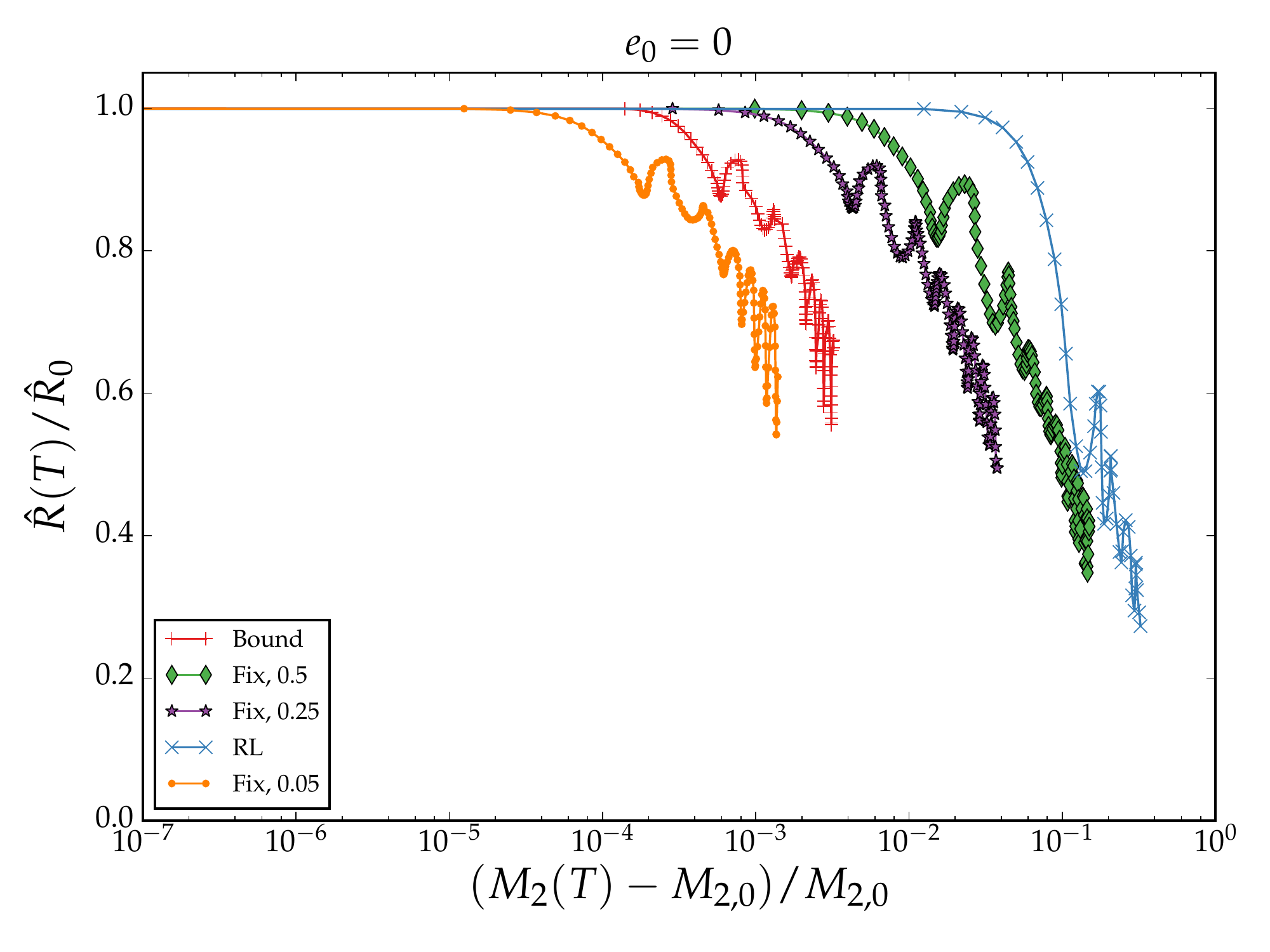}
\caption
{
Binary separation as a function of the secondary mass,
for different accretion prescriptions. The binary separation and
$M_2$ are normalized to their initial values. The evolution is followed
for a time equal to $6\,\Omega_0^{-1}$. Line
colour and style codes are the same as in Figure~\ref{ashort}.
}
\label{avmshort}
\end{figure}

Figure~\ref{avmshort} shows the MBHB separation as a function of $M_2$, the
mass the secondary MBH, to illustrate again that the RL model represents the
less efficient mechanism of binary hardening in terms of normalized accreted
mass. In our case, the high efficiency is caused by the drastically reduced
amount of matter accreted onto $M_2$ (see Figure~\ref{mshort}). The
not-accreted gas still exerts a non negligible gravitational torque onto the
secondary, breaking its orbit and driving its pairing. Such a torque is
incorrectly computed (and overestimated) in the simulations with unphysically
high accretion rates, where the gas responsible for most of the torque is
removed from the simulations and its mass artificially added to the secondary.

\subsection{Evolution of eccentric binaries}
\begin{figure}
    \centering
    \begin{minipage}{.45\textwidth}
      \centering
      \includegraphics[width=\textwidth,clip]{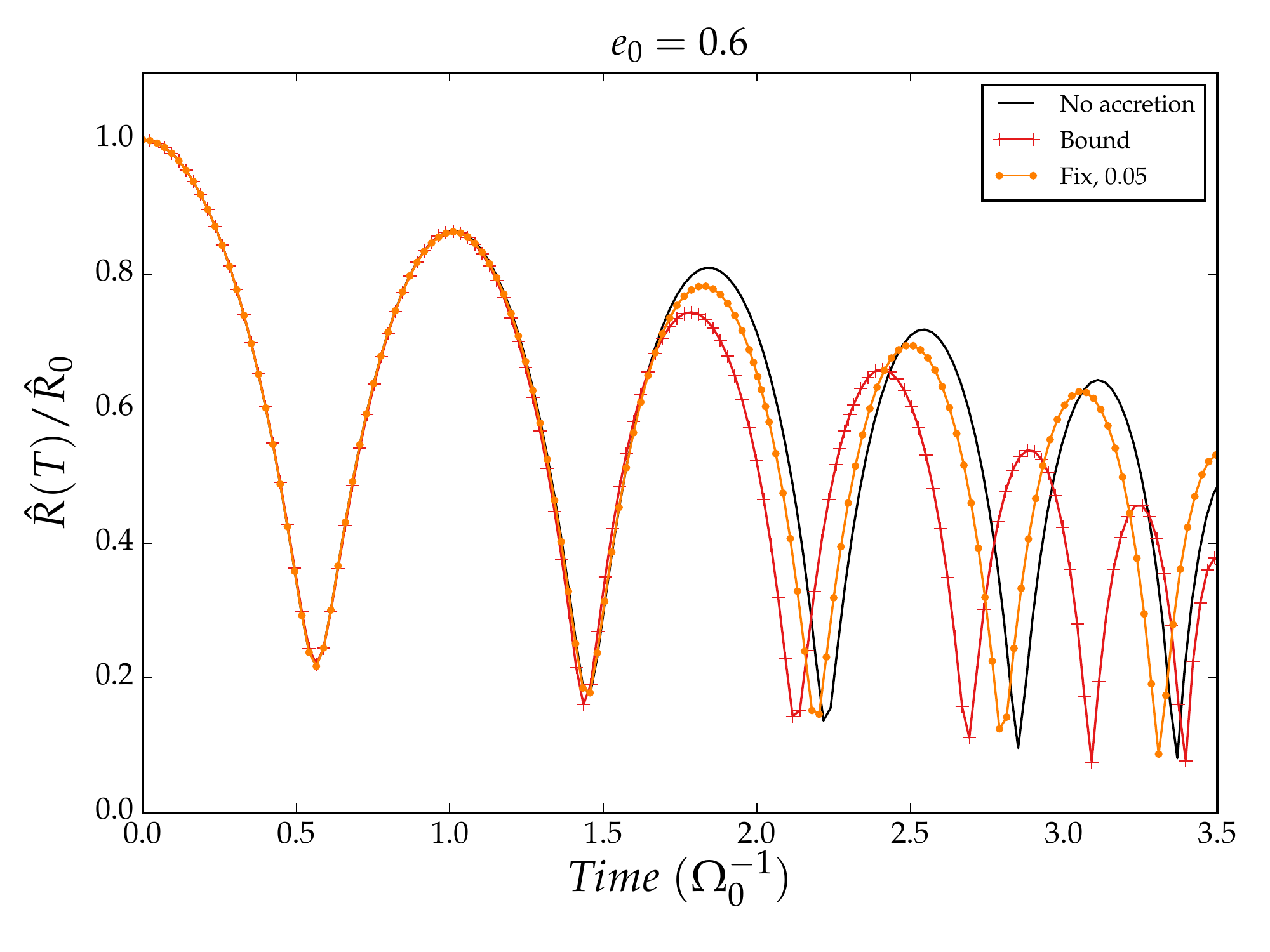}
      \includegraphics[width=\textwidth,clip]{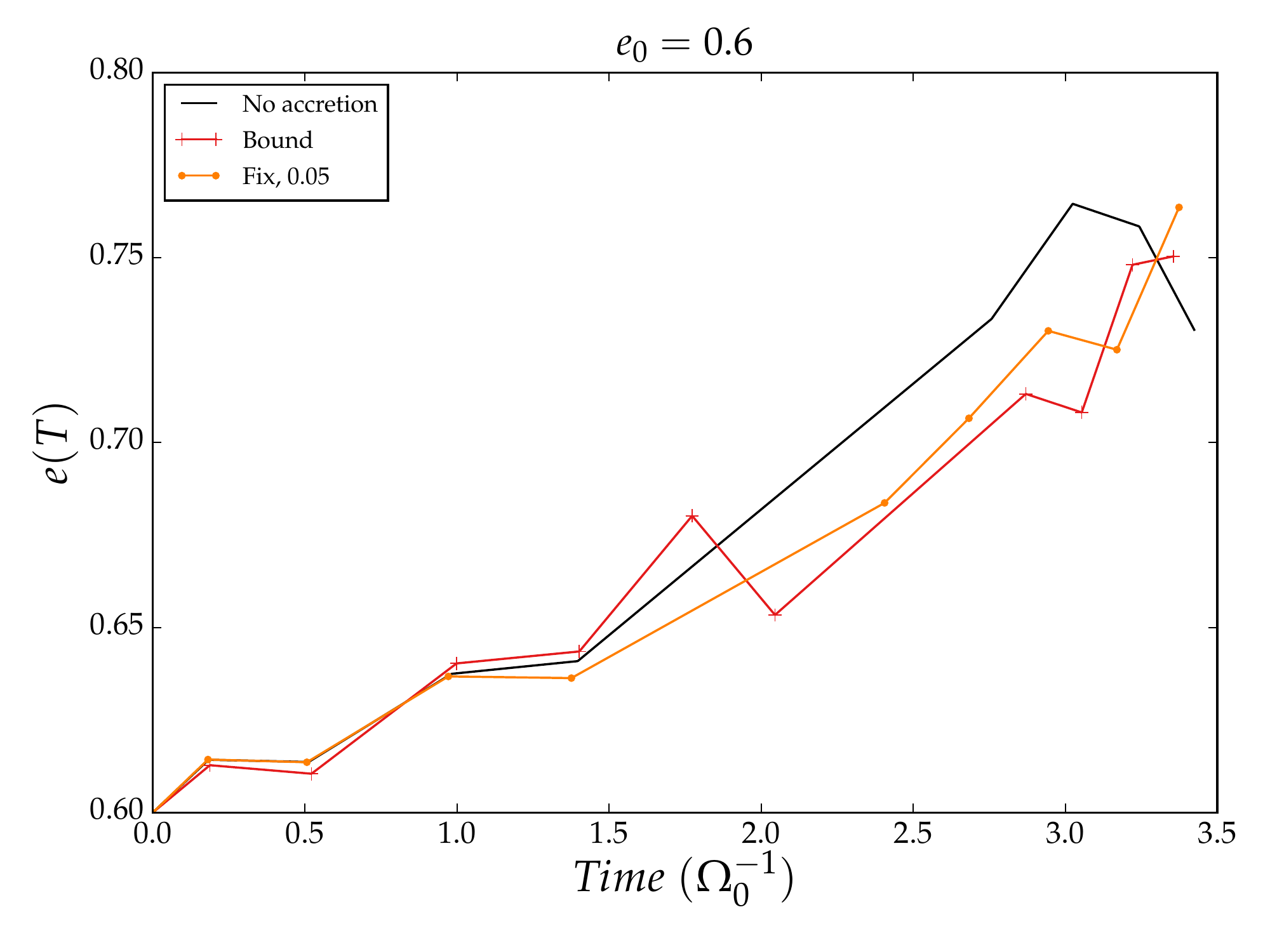}
    \end{minipage}
    \caption{ Upper panel: binary separation versus time,for an eccentric
      binary with $e_0=0.6$.  Units and colour codes are as in the previous
      figures.Lower panel: evolution of the binary eccentricity versus time
      for an eccentric binary with $e_0=0.6$.  }
    \label{fig.Separation_Eccentricity_Ecc0p6_TIME}
\end{figure}

In this section we study the hardening of an initially eccentric binary, in
the retrograde disc.  Because of the results about the accuracy of the
dynamical evolution of the binary presented for the circular case, we confine
the analysis of the eccentric cases to the Bound, No-accretion and $R_{\rm
  fix}=0.05$ runs. The evolution of MBHB separation (upper panel) and
eccentricity (lower panel) for a binary with initial eccentricity $e_0=0.6$ is
shown in figure~\ref{fig.Separation_Eccentricity_Ecc0p6_TIME}.

\begin{figure}
\resizebox{\hsize}{!}
          {\includegraphics[scale=0.5,clip]{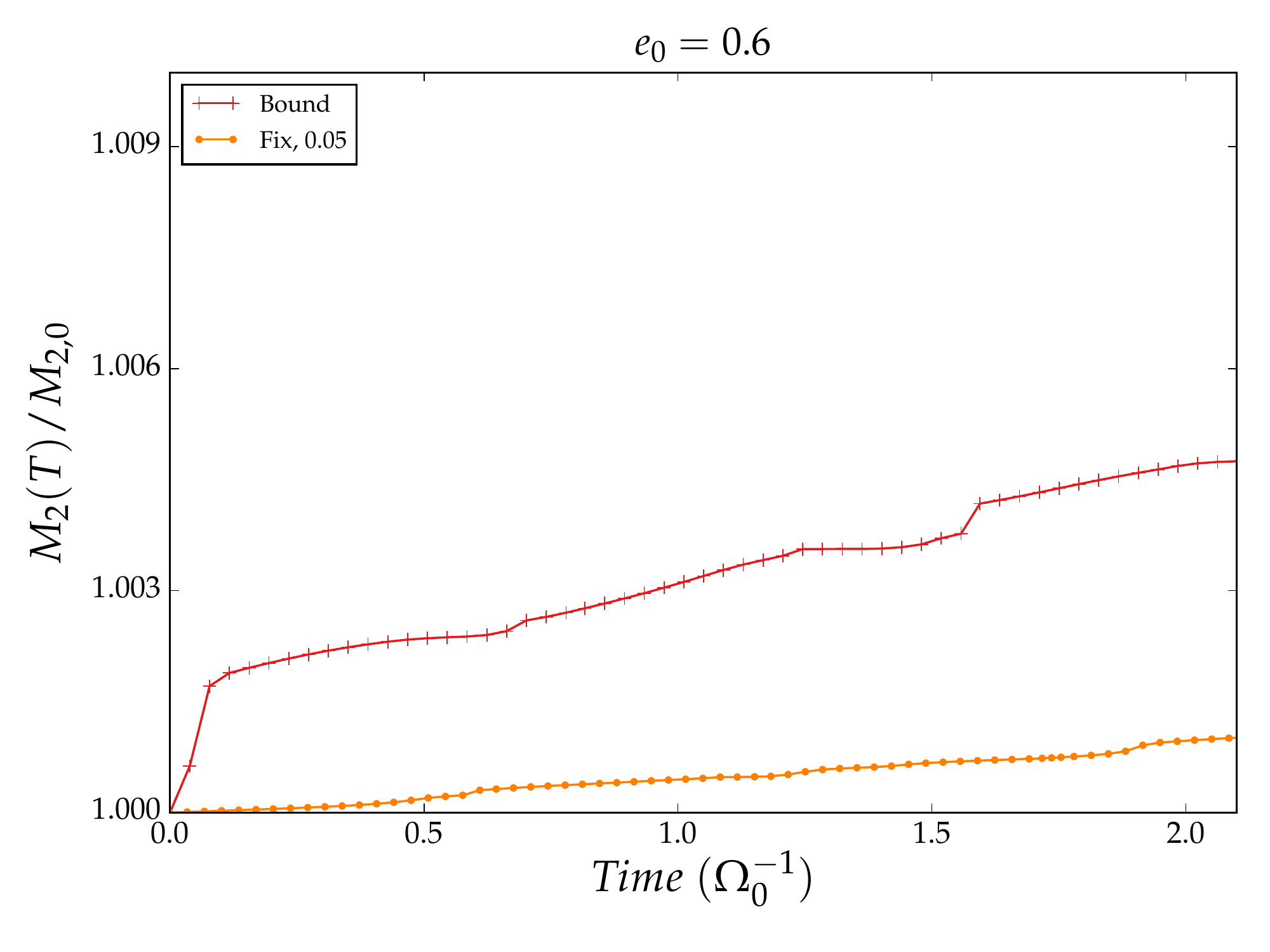}}
\caption
{
    Evolution of the mass of the secondary ($M_2/M_{2,0}$) as a function of time
    (in units of $\Omega_0^{-1}$)  for the eccentric case, with $e_0=0.6$.
}
\label{fig.Mass_Secondary_Retro_Ecc0p6}
\end{figure}

The initial ($t\lsim 2 \Omega_0^{-1}$) evolution of the MBHB is quite similar
in the three cases. The binary hardens and contemporarily the eccentricity
grows considerably, up to $e \gsim 0.7$, in qualitative agreement with the
analytical predictions of~\cite{nixon2011b}. The slight differences between
the different cases are due to the different amount of mass accreted.  As
shown in Figure~\ref{fig.Mass_Secondary_Retro_Ecc0p6} the secondary accretes
more gas when assuming the bound prescription with respect to, e.g., the
$R_{\rm fix}=0.05$ run.

Such differences depend on the accretion prescriptions used.  The Bound and
No-accretion prescriptions avoid implicitly the inconsistency of maintaining a
sink radius $R_{\rm fix}$ constant throughout the orbital phase, over the time
span explored.  In the eccentric case the secondary MBH experiences, along a
single orbit, different regions of the disc. The disc density is the highest
at pericenter where the relative velocity between the gas and the MBH is also
the highest. The notion of $R_{\rm bound}$ becomes thus time (phase)
dependent.  The outcome of the Bound, No-accretion and Fix models can thus be
different.  The run with $R_{\rm fix}$ can give a consistent description of
the MBH dynamics only if, along the orbital phase, $R_{\rm fix}$ remains
smaller than the gravitational influence radius. On the other hand on an
eccentric orbit, the bound prescription over-predicts the amount of matter
accreted.  Gas that is bound at apocenter is accreted promptly when using the
bound prescription. However, gas bound at apocenter would unbind at pericenter
due to the closer action of tidal forces from the primary MBH.  The
instantaneous capture prescription in the Bound model thus over-predicts the
mass accreted, affecting the uderlying dynamics of the disc. In other terms, a
small $R_{\rm fix}$, smaller than the bound radius $R_{\rm bound}$ (at any
time over orbital evolution), or the No-accretion model that allows gas to
remain dynamically active inside the grid, can account for dynamical processes
that can not be captured by the Bound prescription.

\subsection{Three-dimensional SPH experiments}

A clear advantage of using a 2D modelling is that we are able to run many
different cases with a relatively low computational cost. However, the approach upon
which the 2D models rely must be tested\footnote{We note that a recent release
of the code includes the possibility of 3D models, as described in
\url{http://fargo.in2p3.fr/}}. In particular, it is important to asses the
role of disc thickness in affecting the results.

We hence run a few representative cases using full 3D SPH simulations with
\texttt{GADGET-2}~\footnote{\url{http://www.mpa-garching.mpg.de/~volker/gadget/index.html}}
 (Springel 2005).
The gaseous disc is modelled with 2 $\times$ $10^5$ SPH particles,
following an isothermal equation of state but for the possible heating
term associated with the SPH artificial viscosity.
Such viscosity is needed in order to properly recover the occurrence of shocks
in the gas, as it acts when converging flows are in place.
The viscosity term follows a modified Monaghan-Balsara prescription
\citep{MonaghanGingold1983, Balsara1995},
with the viscosity parameter $\alpha$=0.5 and $\beta=2 \times \alpha$.
Gravity is computed on a oct-tree, with close encounters among particles being
softened through a \citep{MonaghanLattanzio1985} spline kernel with the same softening
parameter of $0.1$ for the BHs and gas particles (in internal units).

We implemented a fixed sink radius $R_{\rm fix}$ and a bound radius $R_{\rm
  bound}$ to mimic the \emph{second} and \emph{third} prescriptions presented
in section~\ref{sec.AccrPrescr}. The same $R_{\rm
  fix}$ used in the 2D simulations is introduced here, i.e. $0.05$, $0.25$ and $0.5$. For the
"third" prescription, we select only gas particles within a fix radius of 1
from each MBH, and among these we check which particles are gravitationally
bound to the black hole. In order to do so, we include a parameter $\alpha=0.3$,
similarly to the implementation discussed in \citep{DottiEtAl2007B}. This
$\alpha$ parameter allows us to require that a gas particle is more bound to
the secondary MBH using the Bondi-Hoyle-Lyttleton radius,
\begin{equation}
R_{\rm bound} = \frac{2\, G\, M_{\rm BH}}{v_{\rm bh}^{2} + C_{\rm s}^{2}},
\label{eq.BoundRadius}
\end{equation}
\noindent
where $v_{\rm bh}$ is the relative velocity between the secondary MBH and the
gas particle, and $C_{\rm s}$ is the sound speed of the gas.  We accrete
particles whenever their separations from the MBHs are less or equal to
$\alpha$ times $R_{\rm bound}$.
We initialize the disc using
\texttt{GD\_BASIC}~\footnote{\url{http://www.dfm.uninsubria.it/alupi/software.html}}~\citep{Lupi2014}
following the same prescription for the 2D case, presented in
Section~\ref{sec.NumTool}, preserving the same aspect ratio $H/{\hat R}=0.04$.
\begin{figure}
\resizebox{\hsize}{!}
          {\includegraphics[width=0.8\textwidth,clip]{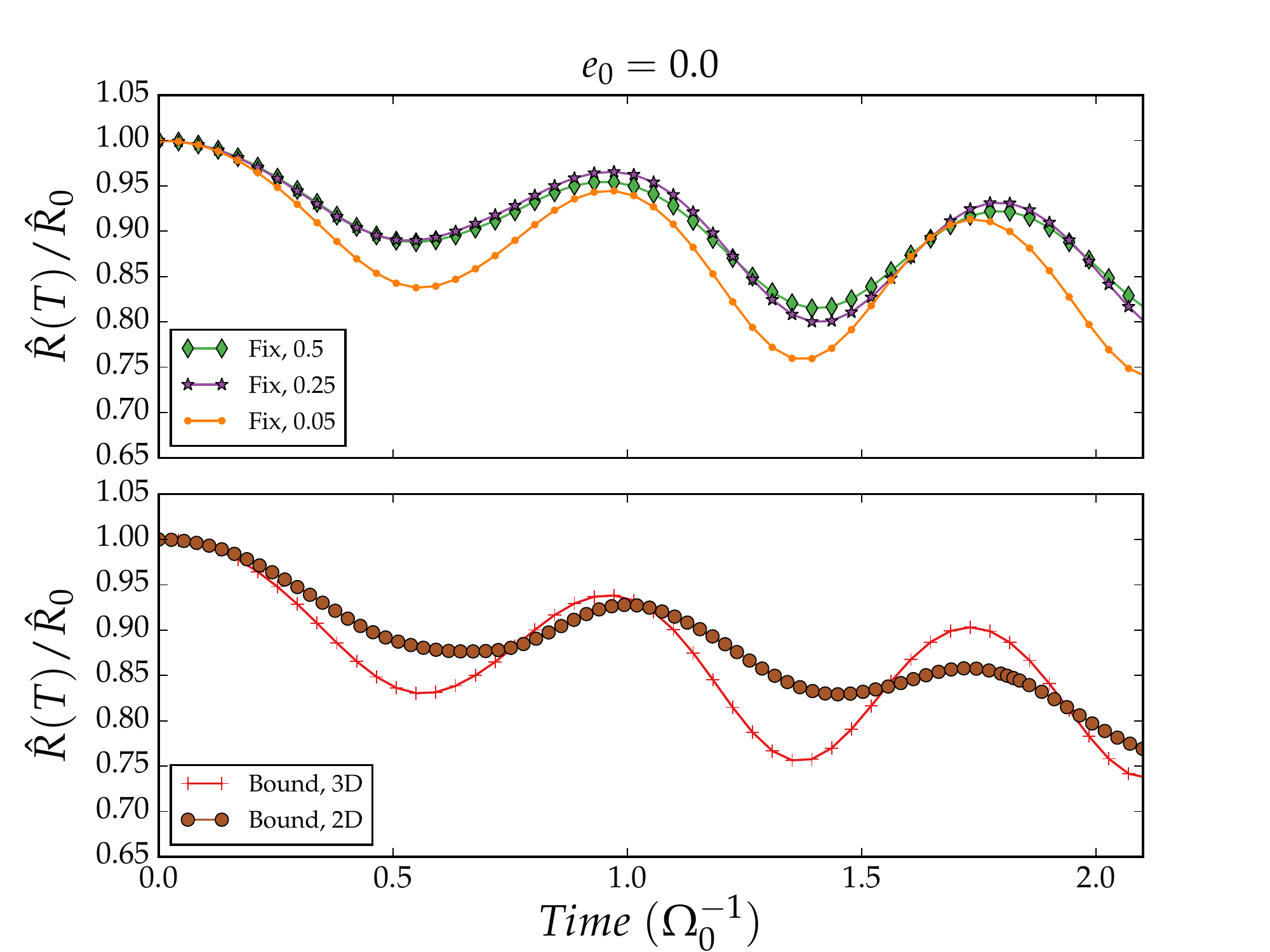}}
\caption
{
Same as Figure~\ref{ashort},
MBHB Orbital separation as a function of time
(in units of $\Omega_0^{-1}$) for the circular case, with $e_0=0$.
{Upper\ Panel:} Evolution of the 3D runs assuming the $R_{\rm fix}$ and $R_{\rm bound}$ accretion prescriptions.
{Lower\ Panel:} Comparison between the $R_{\rm bound}$ prescription
between the 2D and 3D experiments.
}
\label{fig.Separation_Time_Ecc0p0}
\end{figure}
\begin{figure}
\resizebox{\hsize}{!}
          {\includegraphics[width=0.8\textwidth,clip]{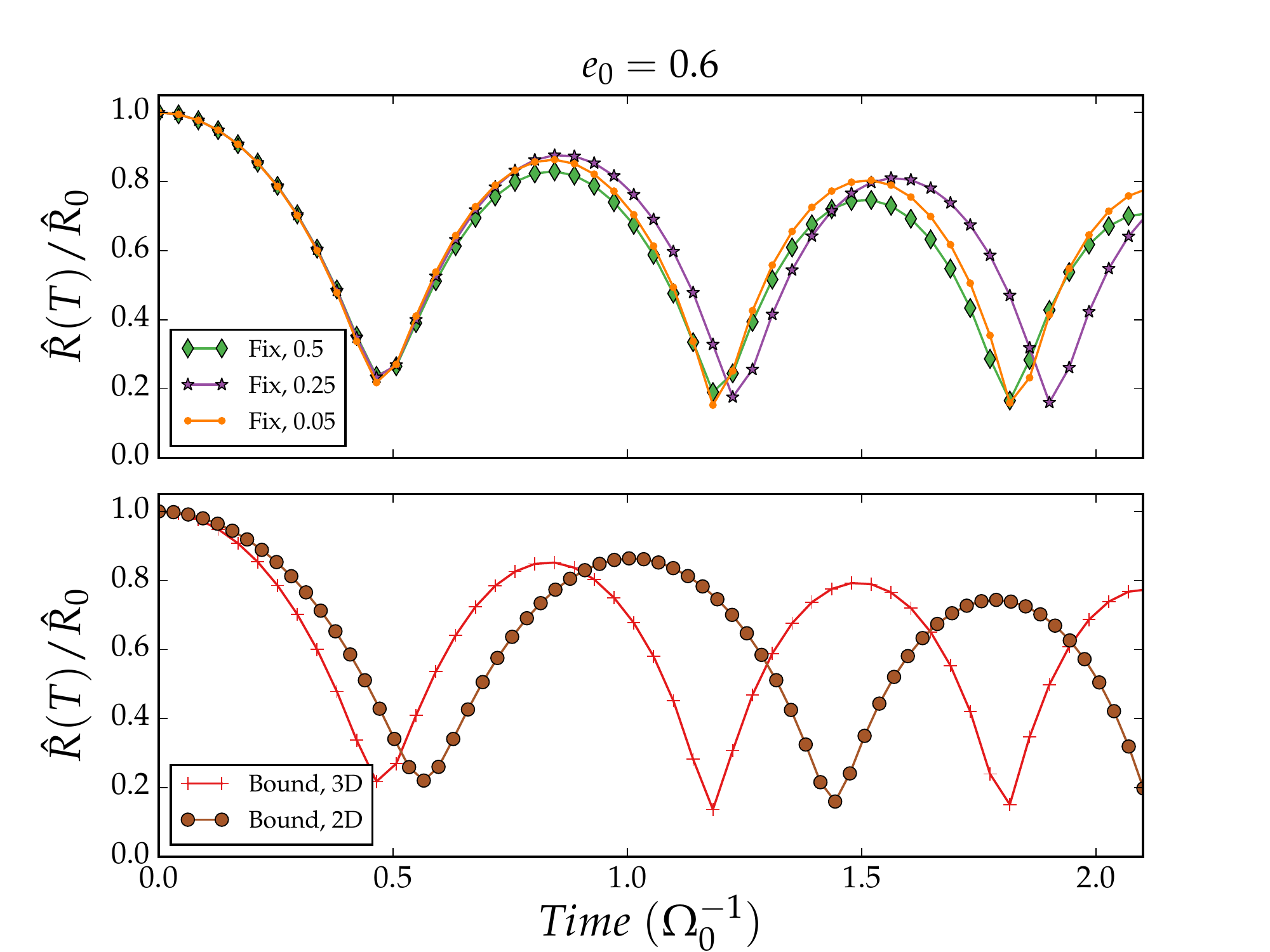}}
\caption
{
Same as Figure~\ref{fig.Separation_Eccentricity_Ecc0p6_TIME}:
orbital separation of the binary as a function of time
(in units of $\Omega_0^{-1}$) for the eccentric case, with $e_0=0.6$.
{Upper\ Panel:} Evolution of the 3D runs assuming the $R_{\rm fix}$ and $R_{\rm bound}$ accretion prescriptions.
{Lower\ Panel:} Comparison between the $R_{\rm bound}$ prescription
between the 2D and 3D experiments.
}
\label{fig.Separation_Time_Ecc0p6}
\end{figure}
We depict the orbital separation between the two MBHs for the
three-dimensional simulations in Figure~\ref{fig.Separation_Time_Ecc0p0} and
\ref{fig.Separation_Time_Ecc0p6} for the circular and eccentric case,
respectively. So as to facilitate the comparison with the 2D case, we have
added a panel in which we display the bound cases for both the 2D and 3D
simulations.  After 5 orbital periods, which require about one week of
computation on 12 CPUs, the 3D experiment reaches a value of $\approx\,0.7$
instead of the $\approx\,0.8$ value of the 2D simulation.
  Two important results can be inferred
  from the comparison of the different runs: \\ $\bullet$ All the discussion
  about the importance of the accretion prescription already presented for the
  2D runs is equally valid in their 3D counterparts. The very similar
  evolution of the MBH separation as a function of time for the $R_{\rm
    fix}=0.05$ and $R_{\rm bound}$ case strengthen the claim that resolving
  the MBH sphere of influence is a necessary requirement for a proper
  dynamical evolution.\\ $\bullet$ We can readily see that the agreement
  between the 2D and 3D runs in Figure~\ref{fig.Separation_Time_Ecc0p0} is
  satisfactory: in spite of the very different numerical approaches and the
  simplifications introduced in the 2D case, the evolution of the distance of
  the binary is very similar, although not identical. The small differences
  present in the MBH evolution are due to the fact that, in the 2D case the
  gas and the MBHs are confined to the plane of the disc and, hence, the MBHs
  are forced to interact with more gas than in the 3D case. In the 3D case,
  however, the non-negligible disc thickness allows some of the gas not to
  interact with the MBH.

\begin{figure}
\resizebox{\hsize}{!}
          {\includegraphics[width=0.8\textwidth,clip]{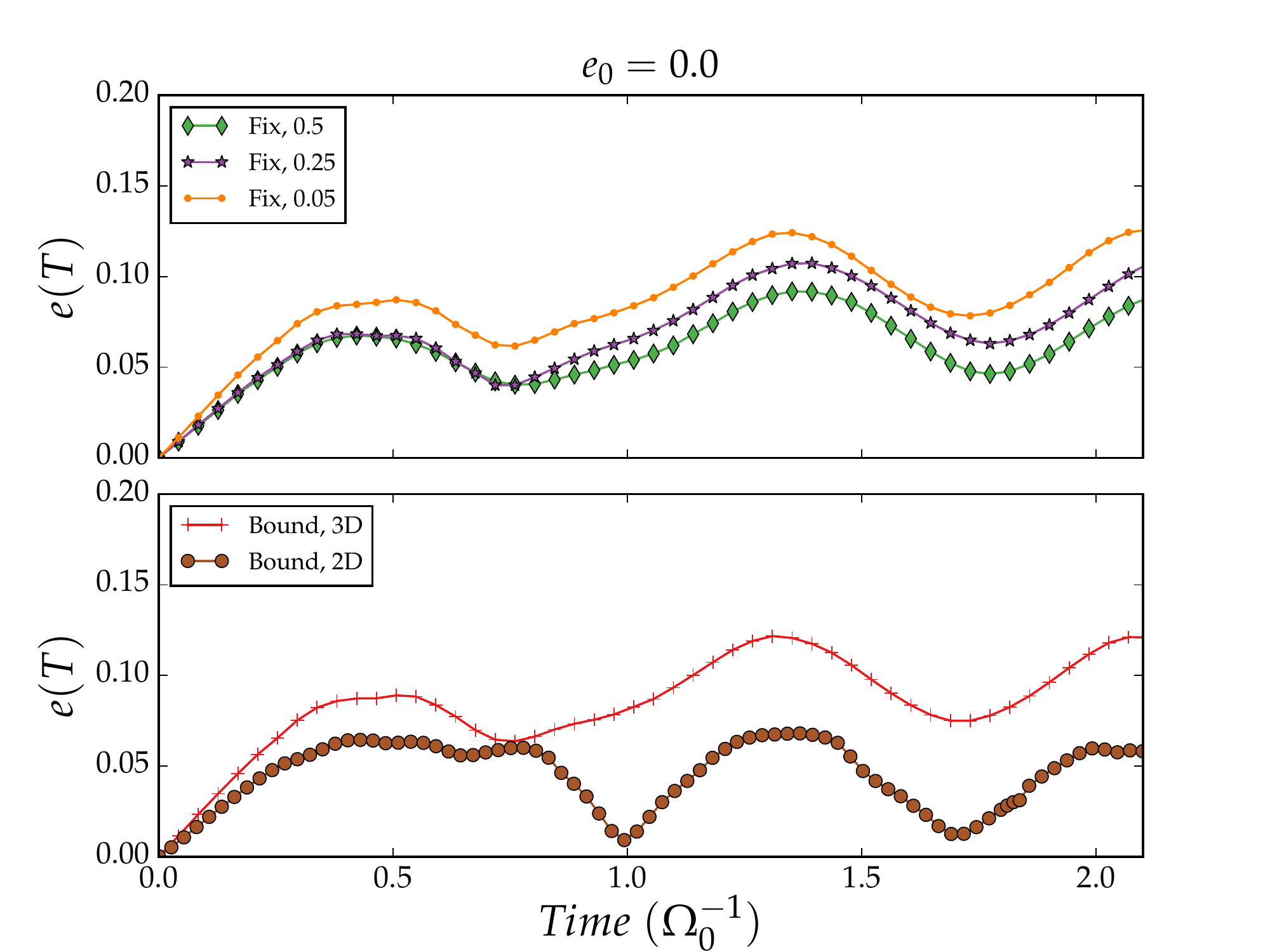}}
\caption
{
Same as Figure~\ref{fig.Separation_Ecc_Time_Circular},
Evolution of the MBHB eccentricity as a function of time
(in units of $\Omega_0^{-1}$) for the circular case, with $e_0=0$.
{Upper\ Panel:} Evolution of the $R_{\rm fix}$ accretion prescriptions.
{Lower\ Panel:} Comparison between the $R_{\rm bound}$ prescription
between the 2D and 3D experiments.
}
\label{fig.Eccentricity_Time_Ecc0p0}
\end{figure}

The effect of the disc vertical profile on the secondary dynamics is
  way more significant in the eccentric case.  In
  Figure~\ref{fig.Separation_Time_Ecc0p6} we see a much more pronounced
  difference between the 2D and 3D bound cases (lower panel).

  As a note of caution we remark that neither the 2D nor the 3D simulations are
  realistic and accurate representations of circum-binary discs. The idealized
  thermodynamics of the gas does not catch all the cooling processes taking
  place in such high density regions. If the disc were allowed to radiate
  a significant amount of the energy acquired from the interaction with the
  MBH, it would settle in a geometrically thinner configuration, more similar
  to the 2D case. In this study we aimed at isolating the effect of the
  accretion prescriptions on the MBH dynamics, and a detailed study of the effect
  of the gas thermodynamics is beyond the scope of the paper.

\begin{figure}
\resizebox{\hsize}{!}
          {\includegraphics[width=0.8\textwidth,clip]{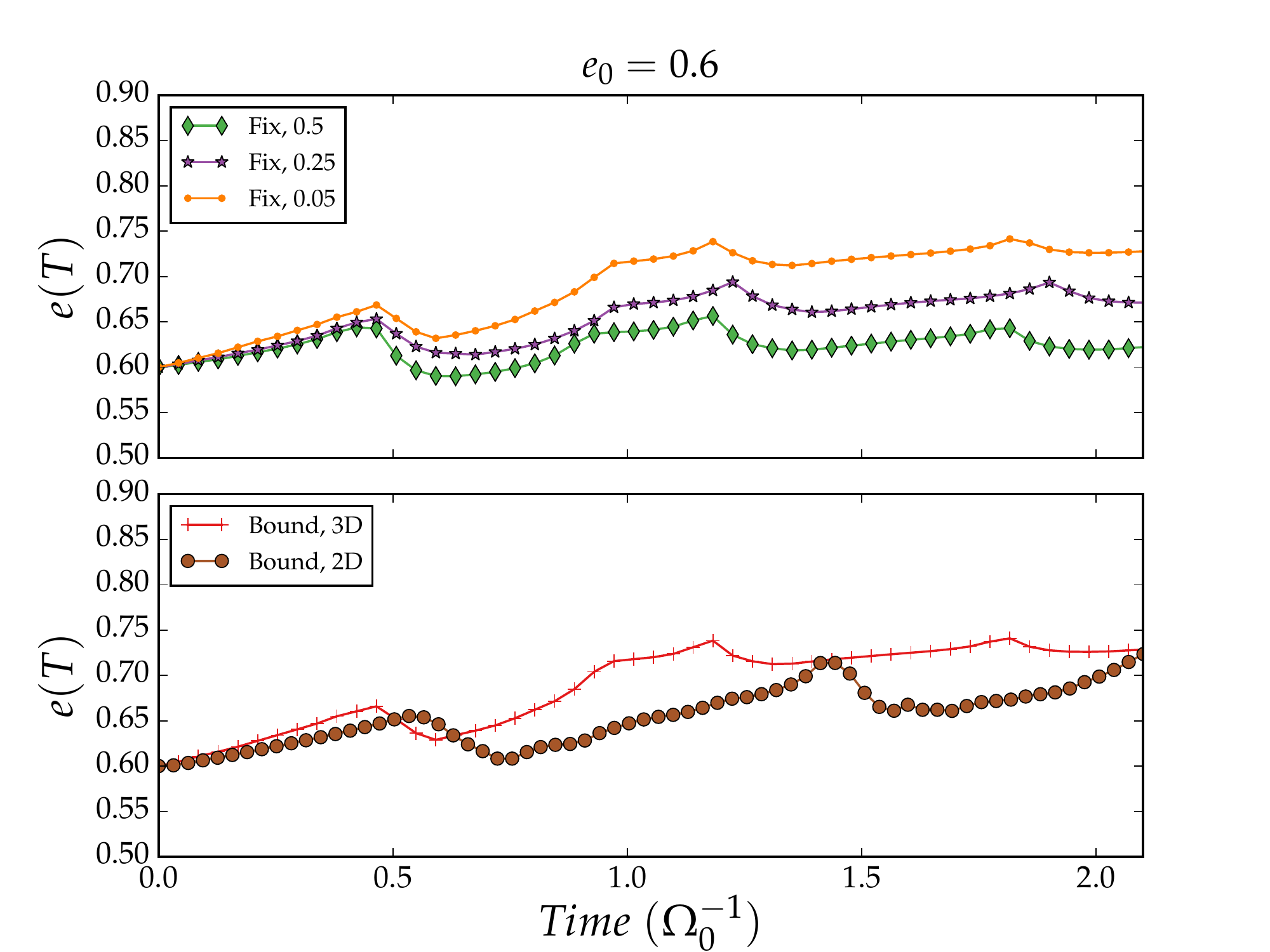}}
\caption
{
Same as Figure~\ref{fig.Separation_Eccentricity_Ecc0p6_TIME},
Evolution of the MBHB eccentricity as a function of time
(in units of $\Omega_0^{-1}$) for the eccentric case, with $e_0=0.6$.
{Upper\ Panel:} Evolution of the $R_{\rm fix}$ accretion prescriptions.
{Lower\ Panel:} Comparison between the $R_{\rm bound}$ prescription
}
\label{fig.Eccentricity_Time_Ecc0p6}
\end{figure}

The comparison of the evolution of the secondary eccentricity in the different
runs gives similar results.  Figure~\ref{fig.Eccentricity_Time_Ecc0p0} shows
the eccentricity as a function of time for the circular cases. The trends are
similar, although not identical in the 2D and 3D bound cases, reaching the
same maximum value ($\approx 0.12$). The differences between these two cases
are due to the reduced amount of gas mass the secondary interacts with in the
3D case, in which the disc thickens as the time goes by.  The comparison among
the 3D runs demonstrates again the need of resolving the secondary sphere of
influence, with the bound and $R_{\rm fix}=0.05$ resulting in the same
dynamical evolution of the binary. The same comments apply to the eccentric
cases shown in Figure~\ref{fig.Eccentricity_Time_Ecc0p6}.

\section{A semi-analytical model for the evolution of a binary in an unperturbed retrograde disc}
\label{sec.Semianal}

In a retrograde disc, the secondary MBH experiences a drag force
resulting from gas-dynamical friction on scales larger than $R_{\rm
  bound}$ and from accretion on scales smaller than $R_{\rm
  bound}$.  Here we propose an analytical scheme that helps interpreting the run of the eccentricity and semi-major axis of the
  MBHB versus time (or equivalently the accreted mass), varying  the slope of the underlying gas density profile.

In the simplifying assumption that the
gaseous background is stationary and that the MBH motion is supersonic,
the deceleration force can be approximated as
\begin{equation}
 {\bf F}_{\rm drag}=-4\pi\lambda G^2M_2^2\rho_{\rm gas}
{{ {\bf V}}_{\rm rel}\over V^3_{\rm rel}},
\label{dragforceacc}
\end{equation}
where $\rho_{\rm gas}$ is the density of the gas at distances near $R_{\rm bound},$
and ${\bf V}_{\rm rel}=V_{\rm rel}{\tilde {\bf V}}_{\rm rel}$ is the velocity of the accreting MBH relative to the gas velocity.
The factor $\lambda$  identifies with the eigenvalue of the Bondi-Hoyle-Lyttleton model for accretion
(equal to $1.12$ for a isothermal gas) but here $\lambda$, suitably rescaled, can also account for the gas
dynamical drag according to \cite{ostriker99}.
 In equation \ref{dragforceacc},
${\bf V}_{\rm rel}={\bf V}_2-{\bf V}_{\rm gas},$  where  ${\bf V}_2$ is the velocity of $M_2$ relative to the center of mass of the binary
(we here consider the limit $M_2\ll M_1$ for simplicity, so that the total mass $M=M_1+M_2$ of the binary is approximated to
$M_1$, and the reduced mass $\mu$ as $M_2$) and ${\bf V}_{\rm gas}$ the Keplerian velocity of the gas,
relative to $M_1$, at the
current position $\hat R$ of $M_2$, i.e. ${{\bf V}_{\rm gas}}=(GM_1/{\hat R})^{1/2}{\tilde {\bf V}}_{\rm gas}$, with ${\tilde {\bf V}}_{\rm gas}$ a unit vector in the direction of ${\bf V}_{\rm gas}$.

To explore the MBHB dynamics and describe the sinking of the secondary MBH in
the retrograde disc, we consider the drag force as a
perturbation on the Keplerian motion of $M_2$ in the gravitational potential
of the primary MBH, $M_1$ \cite{VecchioEtAl1994}.  We then trace the dynamics of
$M_2$ computing the change of the orbital elements, i.e.  the energy and
angular momentum, or equivalently the semi-major axis $a$ and eccentricity
$e$ under the action of the mean drag force,
\begin{equation}
\langle {{\bf F} _{\rm drag}}\rangle_T={(1-e^2)^{3/2}\over 2\pi}
\int_0^{2\pi} {{\bf F}_{\rm drag}(\psi)}{d\psi\over (1+e\cos\psi)^2},
\label{meanforce}
\end{equation}
where $\psi$ the orbital phase, and the mean is over the orbital period.
In this way we separate the instantaneous motion of $M_2$  from the motion averaged over an orbital period: here on we will refer to as secular motion, and secular evolution.

The drag force can be cast in the following form in order to separate the modulus
from the direction, both time (phase) dependent:
\begin{equation}
{{\bf F} _{\rm drag}}=-\xi ({\hat R},V_{\rm rel} )\, ({\tilde  {\bf V}}_2-{\tilde {\bf V}}_{\rm gas})
\label{dragvdot}
\end{equation}
with
\begin{equation}
\xi({\hat R},V_{\rm rel})=4\pi\lambda (GM_2)^2\rho_{\rm gas,0}\left ({{\hat R}_0\over {\hat R}}\right )^n{1\over V_{\rm rel}^2}.
\label{chi1}
\end{equation}
In equation \ref{dragvdot} and \ref{chi1}, the distance $\hat R$ and the velocity vectors are
function of phase $\psi$, along the Keplerian motion, while the constants  ${\hat R}_0$ and $\rho_{\rm gas,0}$ denote a reference radius and density in the retrograde, inhomogeneous disc.

 In equation \ref{chi1}, the power-law exponent $n$ describes the
 distribution of the gas density
 as a function of the distance $\hat R$: $\rho_{\rm gas}({\hat R})= \rho_{\rm gas,0}({\hat R}_0/{\hat R})^n$. The value $n=1$
corresponds to the disc's density profile at the onset of the hydrodynamical simulations.

 After some calculation (sketched in Appendix) the equations for the secular evolution of the MBH mass
 $m_2$, in units of $M_{2,0}$, of the semi-major axis
${\tilde a}=a /a_0$,  in units of the initial semi-major axis $a_0$, and of the eccentricity $e$
can be cast in a simple form:
\begin{equation}
{{\dot m_2}\over m_2}=  \Gamma_0 (1-e^2)^{(3-n)} {\tilde a}^{(3/2-n)}\langle {{\cal A}(e)}\rangle_\psi,
\label{dotm}
\end{equation}
where dot denotes
the ``secular'' time derivative, $m_2=M_2/M_{2,0}$, and $\langle {{\cal A}(e)}\rangle_\psi$  a dimensionless function of the running value of  $e$ (eq. in Appendix).  $\Gamma_0$ is a constant equal to

\begin{equation}
 \Gamma_0=4\pi\lambda G\rho_{\rm gas,0}{1\over \Omega_0}{M_{2,0}\over M_1}\left ({{\hat R}_0\over a_0}\right )^n.
 \label{gamma}
\end {equation}
In this last expression $\Omega_0$ is the Keplerian frequency of the MBHB at $a_0:$ $\Omega_0=(GM_1/a_0^3)^{1/2}$.
The equation for the dimensionless semi-major axis reads

\begin{equation}
{{\dot {\tilde a}}\over {\tilde a}}=-2\,\, \Gamma_0\,m_2\,{\tilde a}^{(3/2-n)}\,[1-e^2]^{(2-n)}\langle {{\cal B}(e)}\rangle_\psi,
\label{dota}
\end{equation}
 where $\langle {{\cal B}(e)}\rangle_\psi$ is given in Appendix.

Similarly, one can calculate the rate of change of the angular
momentum and in turn of the eccentricity $e$.  As the orbit of $M_2$
is coplanar, the direction of the
binary orbital angular momentum does not vary with time as the drag is
causing only a decrease in the modulus of {$\bf J$}.  According to
equation \ref{dragvdot}, the eccentricity evolves as
\begin{equation}
{{\dot e}\over e}=\Gamma_0\,\,m_2\, {\tilde a}^{(3/2-n)}
\,\, {(1-e^2)^{(4-n)}\over e^2}\,\, \langle {{\cal C}(e)}\rangle_\psi,
\label{dote}
\end{equation}
where $ \langle {{\cal C}(e)}\rangle_\psi$ is a dimensionless function of the running  value of the eccentricity $e$ (eq. in Appendix).

Equations \ref{dotm}, \ref{dota} and \ref{dote} are coupled and can be solved numerically for $m_{2,0}=1$, ${\tilde a}_0=1$ and initial eccentricity $e_0$ at time $t=0$.
The  results can then be rescaled for any arbitrary value of $M_{2,0}$ and $a_0$.
In \ref{dotm}, \ref{dota} and \ref{dote}, the timescale that enters the equations  is   $\tau_0=\Gamma_0^{-1}$
that can be displayed in the form
\begin{equation}
\tau_0\sim
{1\over 8\lambda}{V_{\rm K,0}\over \pi G \Sigma_{\rm gas,0}} {H\over {\hat R}_0}{M_1\over M_{2,0}}\left ({a_0\over {\hat R}_0}\right )^n,
\end{equation}
where $V_{\rm K,0}$ is the Keplerian velocity at ${\hat R}_0$, and $\Sigma_{\rm gas,0}\sim 2H\rho_{\rm gas,0}$.
Recalling that $H/{\hat R}_0\sim c_{\rm s,0}/V_{\rm K,0}$ in a thin isothermal disc
(with isothermal sound speed $c_{\rm s,0}$),
$\tau_0$ scales as
\begin{equation}
\tau_0\Omega_0\sim
{1\over 8\lambda}Q_0 {M_1\over M_{2,0}},
\end{equation}
where we have introduced the Toomre parameter $Q_0=c_{\rm s,0}\Omega_0/(\pi G\Sigma_{\rm gas,0})$ for disc stability.
In the following we will describe the solutions of this simple model in the $e$ versus
$M_2/M_{2,0}$ and $\tilde a$ versus  $M_2/M_{2,0}$  planes.

\subsection{Binary evolution trends}

In a fixed and non steep gaseous background  ($n<3$), eccentric binaries evolves into more eccentric binaries
\footnote{We notice that a rapid increase in the eccentricity is also found in an analytical study by \cite{Schnittman15} who expressed the
negative torque of a retrograde disc on a
MBHB as function of the mass accretion rate, motivated by magneto-hydrodynamical
simulations by Bankert et al. (2014).} .

\begin{figure}
\resizebox{\hsize}{!}
          {\includegraphics[scale=1,clip]{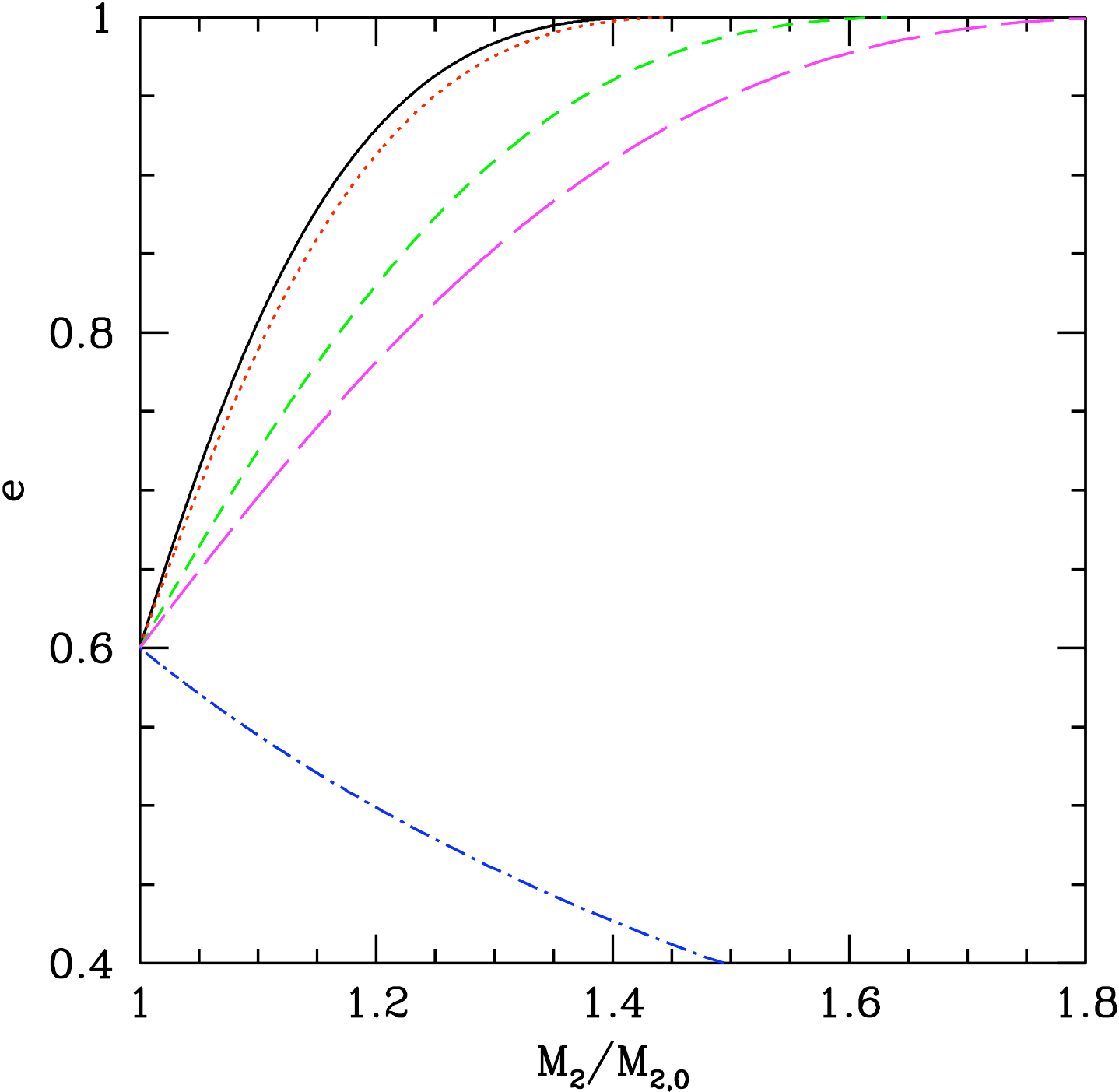}}
\caption
   { Orbital eccentricity $e$ versus $M_2/M_{2,0}$ for $e_0=0.6$, as computed
     within the semi-analytical model.  Solid (black) line refers to a
     background gas density scaling with distance $\hat R$ as a power law with
     $n=1,$ corresponding to the profile of the hydrodynamical simulation. Red
     (dotted), green (dashed), magenta (long dashed) and blue (dot-dashed)
     refer to retrograde discs with $n=1.3,2,2.2$ and $3$, respectively. }
\label{fig.eccmass}
\end{figure}

Figure \ref{fig.eccmass} shows the run of the eccentricity $e$ versus
$M_2/M_{2,0}$ for $e_0=0.6$, and $n=1,1.3,2,2.2$ and 3. The
eccentricity grows monotonically up to unity, and this occurs before $\tilde
{a}$ has decayed significantly (i.e., by more than two orders of magnitude). The limit $e\to
1$ is reached after the secondary has accreted a mass comparable to
$30\%-60\%$ of the initial
mass $M_{2,0}$.

In the case of a steep density background, i.e. $n=3$,
the eccentricity shows an opposite trend and decreases with time,
since the gas disc density at
pericenter is so high that the secondary MBH experiences the largest drag there.
A higher braking force at pericenter reduces the eccentricity and the MBH
spirals inwards along orbits progressively less eccentric. The semi-major axis
drops dramatically by more then three orders of magnitude, on a time $\tau_0$.

Nearly circular orbits (with $e_0=0.01$) show interesting behaviours in
their long-term evolution.
We first notice that $n=3/2$ is a critical value,
separating two trends. Solutions with $n<3/2$ evolve toward
$e\to 1$ before the semi-major axis has
decreased significantly. By contrast, when $n>3/2$, the decay of the
semi-major axis is faster than the increase of $e$.  The value of
$n=3/2$ is critical since the logarithmic derivatives of our variables all scale as
${\tilde a}^{(3/2-n)}$, as indicated in equations \ref{dotm},
\ref{dota}, and \ref{dote}. For $n>3/2$ the decay of the semi-major axis
accelerates with decreasing  $\tilde a$.

\begin{figure*}
\resizebox{\hsize}{!}
   {\includegraphics[scale=1,clip]{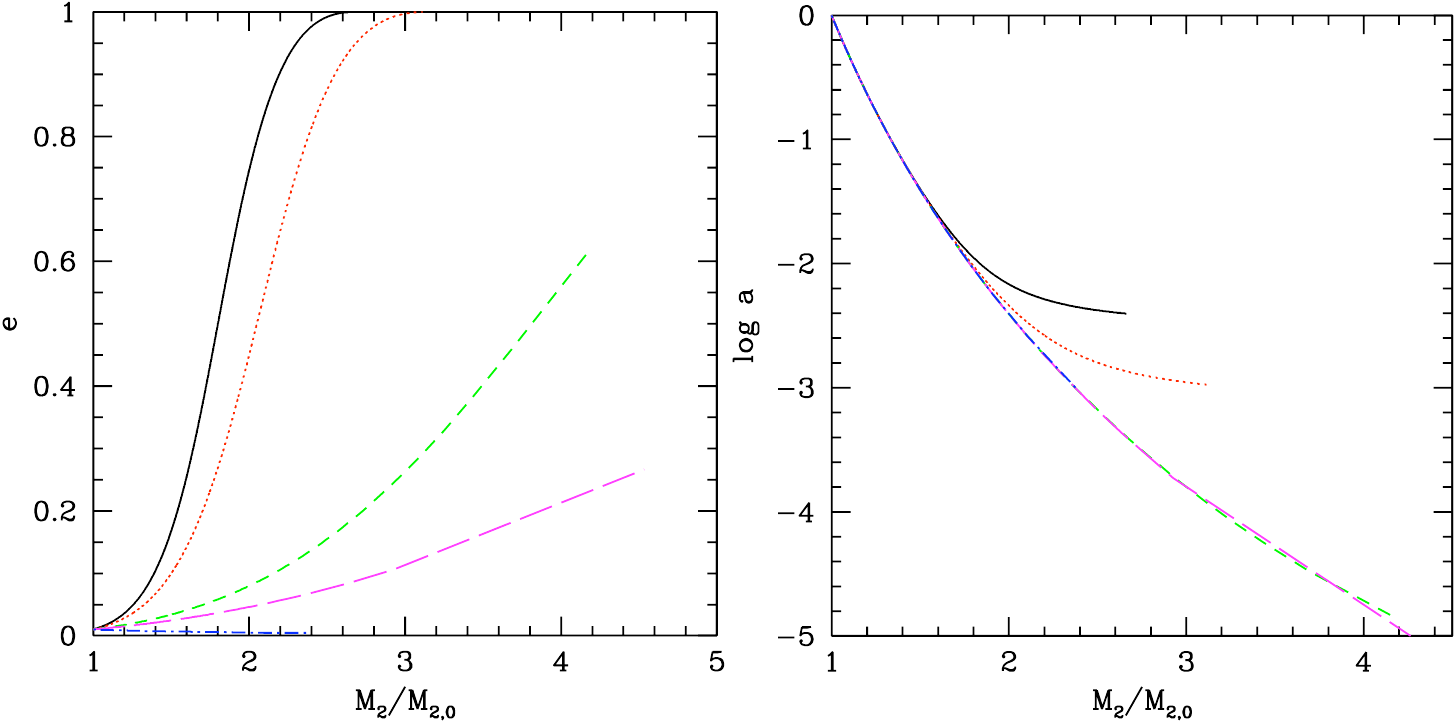}}
        \caption
   { Eccentricity $e$ (left panel) and semi-major axis $\tilde a$ (right
     panel) versus $M_2/M_{2,0}$ for a nearly circular orbit, with $e_0=0.01$,
     as computed within the semi-analytical model. Line colors and style codes
     are the same as in Figure \ref{fig.eccmass}.  For a retrograde disc with
     background density $\rho_{\rm gas}\propto {\hat R}^{-n}$ scaling with
     $n<3/2$ (solid-black, red-dotted lines) the evolution stops at the time
     the eccentricity $e\to 1$. By contrast, when $n>3/2$ (as in the remaining
     cases, i.e., $n=2,2.2$ and $3$) the semi-major axis decays very rapidly
     before the eccentricity has time to rise up to unity. }
\label{fig.emam001}
\end{figure*}

Figure \ref{fig.emam001} shows  the run of $e$ and $\tilde a$ versus
$M_2/M_{2,0}.$
As a rule of thumb and for circular binaries, the
secondary needs to accrete a mass of the order of (2-4) $M_{2,0}$ to
slide on the path leading to coalescence by gravitational waves.
For $n=3$,
the eccentricity remains small for the entire evolution, and never rises.

In the early phase of the binary evolution (corresponding to
$M_2<0.2M_{2,0}$), nearly circular binaries remains almost circular,
as their eccentricity growth occurs after a sizable increase of the MBH mass (see Fig.~\ref{fig.emam001}).
This explains why the initially
circular binary models studied with FARGO with different accretion
prescription do not display a sizable increase of $e$, on the time
span of the simulation. This is in line with the findings by \cite{roedig14}.

Our semi-analytical model has applicability in the limit of $M_{2}/M_1\ll 1$.
Only if  $M_{2}/M_1\ll 1$
the secondary, lighter MBH causes minor perturbations
in the underlying disc  so that the background density can be treated as a constant
over the binary  evolution timescale. For larger MBH ratios (as the one considered in this paper) changes in
the density background occur in response to the mutual
perturbation excited by the secondary MBH in the retrograde disc, and
the evolution needs to be traced only via direct numerical
simulations.

\section{Discussion}
\label{sec.Discuss}

In this paper we explored the evolution of a MBHB embedded in a counter-rotating
circum-binary disc.  As mentioned in the introduction, the retrograde
case differs from the prograde one in three main points: (i) the
gravitational torque responsible for the binary shrinking does not
halt  inflows of gas around the MBHs embedded in the disc;  (ii) retrograde gas interacting with
the MBHs can remove more angular momentum per unit of mass, since its
initial angular momentum has  sign opposite to that of the binary;
(iii) in the retrograde case, the relative velocities between the MBHs
and the disc are significantly larger, so that the interaction between
the gas and the MBHs is limited to smaller regions. Since points (i) and
(ii) facilitate the binary shrinking while point (iii)  limits its
strength, it is not obvious a priori if the interaction between a MBHB and a
counter-rotating circum-binary disc results in rapid hardening of the binary.

In particular, we pointed out that, as a consequence of the high
relative speed between the secondary and the gas,
simulations of counter-rotating MBHB-disc systems are strongly
affected by the prescription assumed for the accretion of gas onto the
MBHs. We argued that in order to
  model the disc-MBH hydro-dynamics in an appropriate way, a necessary
condition  is that only gas {\it bound} accretes onto the lighter,
secondary MBH. We confirmed our claim running a suite of 2D
hydrodynamical simulations, showing that assuming a too large
accretion (sink) radius, such as e.g. the Roche Lobe radius of the
secondary (\cite{nixon2011b}), results in a too fast spurious evolution of the binary (see
Figures~\ref{ashort},~\ref{mshort},~\ref{avmshort}). We further
noticed that, while for secondaries moving on quasi-circular orbits
the "Bound" prescription results in the correct dynamical evolution of
the secondary, for MBHs moving on very eccentric orbits this
prescription is not sufficient. This is due to the fact that gas bound
to the secondary at the apocenter can unbind during the close
pericenter passages, avoiding to be accreted. The non-accreted gas
changes the effective mass of the secondary (i.e. the mass of the
secondary plus the mass of the gas co-moving to it), and can rejoin
the background gas distribution, possibly further interacting with the
secondary at later times. For this reason, a sink radius smaller than
the influence radius of the MBH, varying over the orbital phase,
has to be set properly in order to predict the binary dynamical
evolution (see Figure~\ref{fig.Separation_Eccentricity_Ecc0p6_TIME},
\ref{fig.Mass_Secondary_Retro_Ecc0p6}).

The 3D SPH simulations run in parallel for selected cases
have shown a close match with the results of the 2D simulations validating
the findings in 2D, despite the intrinsic physical difference
related to the different dimensionality and
the difference in the numerical method.

We focused on the evolution of the binary orbit giving particular emphasis
to the effects of the accretion prescriptions.
The present
analysis still lacks of some possibly important physical
effects.  We limited the mass of the disc to its
initial value, and we do not study the effect of a continuous (or
episodic) re-fueling of the disc from the outer regions of the
nucleus. Such a fueling is necessary for the coalescence of a binary
in a prograde disc~\citep[as discussed in][]{Dotti12}, and can boost
the brake of the binary in the retrograde case as well. Finally, we
did not model any disc fragmentation and star-formation in the disc,
that (together with the fueling of the binary from large scales) is
the main unknown in the evolution of binaries in prograde discs as
well~\citep[see e.g.][]{Lodato09,Amaro-SeoaneBremCuadra2013}. The
effects of these processes, together with a detailed discussion of the
peculiar observational properties of counter-rotating systems, is
postponed to future investigations.

\section{Appendix}
In this Appendix we shortly outline the key steps for deriving the evolution equations
\ref{dotm}, \ref{dota} and \ref{dote}  of Section 5.
We follow closely the derivation by \citep{VecchioEtAl1994}.

First we specify the instantaneous motion of the
secondary MBH, described by the velocity vector ${\bf V}_2$
and the separation vector $\bf {\hat  R}_2$, relative to the center of mass of the binary.  Since the accretion drag
is a weak perturbing force, the motion is determined by the driving force by the MBH primary.
The instantaneous values of $ \bf{\hat  R}_2$
and ${\bf V}_2 $
are expressed in term of the orbital phase $\psi$, along the Keplerian motion. They are
defined uniquely by the instantaneous values of the energy and  angular momentum per unit mass
that we here denote as $E$ and $\bf J$, or alternatively $a$ and $e$.
For the distance modulus we have

\begin{equation}
{{\hat R}}_2=\frac{a(1-e^2)}{1+e\cos\psi}\,.
\label{raggio}
\end{equation}
The velocity is decomposed along the radial and tangential directions
\begin{equation}
V_{2,{\hat R}_2} =\left[\frac{GM}{a(1-e^2)}\right]^{1/2}
e\sin\psi\,,
\label{vr}
\end{equation}

\begin{equation}
V_{2,t}=\left[\frac{GM}{a(1-e^2)}\right]^{1/2}
(1+e\cos\psi)\,,
\label{vt}
\end{equation}
where $M=M_1+M_2\sim M_1$ in the limit of a massive primary.
Furthermore,
\begin{equation}
V_2 = \frac{GM_1}{a(1-e^2)}(1+2 e \cos\psi+e^2)\,.
\end{equation}
The instantaneous deceleration on the secondary MBH due to accretion can be written as

\begin{equation}
{{\dot {\bf V}} _{\rm drag}}=-{1\over M_2} \xi ({\hat R}_2,V_{\rm rel} )\, ({\tilde  {\bf V}}_2-{\tilde {\bf V}}_{\rm gas})
\label{dragacc}
\end{equation}
with
\begin{equation}
\xi({\hat R}_2,V_{\rm rel})=4\pi\lambda (GM_2)^2\rho_{\rm gas,0}\left ({{\hat R}_0\over {\hat R}_2}\right )^n{1\over V_{\rm rel}^2}.
\label{chi}
\end{equation}

The Keplerian velocity of the fluid elements at distance $\hat R$ is ${\bf V}_{\rm gas}=(GM_1/{\hat R})^{1/2}{\bf {\tilde  V}}_{t,\rm gas}$
which is vector tangential to ${\bf {\hat R}}/\hat R$, as the gas in the retrograde disc is moving on circular orbits.
According to equations \ref{raggio}, \ref{vr}, \ref{vt} and \ref{dragacc},
\begin{equation}
{\bf V}_2\cdot {\bf V}_{\rm gas}=-{GM_1\over a(1-e^2)}(1+e\cos\psi)^{3/2}.
\end{equation}

The instantaneous rate of change of the energy per unit mass and angular momentum
per unit mass (in the direction of
$\bf J$, as the orbit is coplanar with the disc's plane )
read:

\begin{equation}
\dot{E}= \dot{\bf {V}}_{\rm drag} \cdot {\bf {V}_2}
\label{Energydot}
\end{equation}
\begin{equation}
\dot{J} =( {\bf {\hat R}}\times \dot {\bf {V}}_{{\rm drag}} )
\cdot { {\bf  {J}}\over { J } }
\label{Jdot}
\end{equation}

The rate of change of the energy per unit mass correlates with the rate of change of the semi-major axis through the relation
${\dot E}/E=-{\dot a}/a$,
where $E=-GM_1/2a$. Similarly, the rate of change of the eccentricity $e$
scales as $e {\dot e}=-(1-e^2)[{\dot J}/J+(1/2){\dot E}/E]$,
where $J^2=GM_1a(1-e^2)$ \citep{VecchioEtAl1994}.
As the secondary MBH accretes gas, the instantaneous rate of
change of the mass $M_2$ is set equal to the Bondi-Holye-Littleton accretion rate:
\begin {equation}
{\dot M}_2={\dot M}_{\rm BHL}.
\label{dotm2}
\end{equation}
The drag force due to accretion induces secular changes in the orbital elements of the binary.
We thus calculate the secular rate of change of $a,e$ and $M_2$
 averaging all physical quantities over the Keplerian period $P.$
As along a Keplerian orbit, the phase $\psi$ is related to the time coordinate by
\begin{equation}
{1\over P}dt={1\over 2\pi}(1-e^2)^{3/2}{1\over (1+e\cos\psi)^2}d\psi,
\label{mean}
\end{equation}
we convert
the means in the time domain to the phase domain,
defining $\langle \cdot\rangle_T=(1-e^2)^{3/2}\langle\cdot(1+e\cos\psi)^{-2}\rangle_\psi$, hereon.

Using the above relations, we derive equations \ref{dotm}, \ref{dota} and \ref{dote}.
If we define $f=1+e^2+2e\cos\psi$ and $g=1+e\cos\psi,$ the expressions for the means  in the phase domain
introduced in equation  \ref{dotm}, \ref{dota} and \ref{dote}  read:

\begin{equation}
\langle {{\cal A}(e)}\rangle_\psi={1\over 2 \pi}\int_0^{2\pi} d\psi{
g^{(n-2)}\over (f+g+2g^{3/2})^{3/2}}
\end{equation}

\begin{equation}
\langle {{\cal B}(e)}\rangle_\psi={1\over2 \pi}\int_0^{2 \pi} d\psi{
g^{(n-2)} (f^{1/2}+g)\over (f+g+2g^{3/2})}
\end{equation}

\begin{equation}
\langle {{\cal C}(e)}\rangle_\psi=
{1\over2 \pi}\int_0^{2 \pi} d\psi{
g^{(n-2)} \over (f+g+2g^{3/2})}{\cal I}(\psi,e)
\end{equation}
where
\begin{equation}
{\cal I}(\psi,e)= \left (f^{-1/2}+g^{-1}-f^{1/2}/(1-e^2)-g/(1-e^2)\right ).
\end{equation}

Notice that in the limit of $e\to 0$, ${\dot e}$ is order ${\cal O}(e)$.
The expansion analysis around $e\to 0$ shows that
${\dot e}/e$ is proportional to $(11/4-n)$.
Thus, for small initial eccentricities and  $n>11/4$, the orbit
circularizes while in the opposite case ($n<11/4$) , $e$ grows in time.
In the same limit ${\dot a}$ is finite, and the MBH continues to spiral in.

\begin{acknowledgements}

It is a pleasure to thank Fr{\'e}d{\'e}ric Masset for his answering questions
regarding {\sc Fargo}, as well as Pablo Ben{\'i}tez Llambay, Jules Casoli and
Aur{\'e}lien Crida.

\end{acknowledgements}

\label{lastpage}
\end{document}